\documentclass[fleqn,usenatbib]{mnras}

\usepackage{newtxtext,newtxmath}
\usepackage[T1]{fontenc}
\usepackage{ae,aecompl}

\usepackage{graphicx}	% Including figure files
\usepackage{amsmath}	% Advanced maths commands
\usepackage{subcaption}
\usepackage{float}
\usepackage{flexisym}
\usepackage{siunitx} % For \num{}

\newcommand{\sus}[2]{\substack{+#1 \\ -#2}}

\title{Optical photometry of two transitional millisecond pulsars in the radio pulsar state}

\author[J. G. Stringer et al.]{J. G. Stringer$^{1}$,
R. P. Breton$^{1}$,
C. J. Clark$^{1,7,8}$,
G. Voisin$^{1,2}$,
M. R. Kennedy$^{1}$,
\newauthor
D. Mata S\'{a}nchez$^{1,3,9}$,
T. Shahbaz$^{3}$,
V. S. Dhillon$^{4}$,
M. van Kerkwijk$^{5}$,
T. R. Marsh$^{6}$ 
\\
$^{1}$Jodrell Bank Centre for Astrophysics, Department of Physics and Astronomy, The University of Manchester, M19 9PL, UK\\
$^{2}$ LUTH, Observatoire de Paris, PSL Research University, 5 Place Jules Janssen, 92195 Meudon, France\\
$^{3}$ Instituto de Astrof\'{i}sica de Canarias, V\'{i}a L\'{a}ctea, E-38205 La Laguna, Tenerife, Espa\~{n}a\\
$^{4}$ Department of Physics and Astronomy, The University of Sheffield, Western Bank, Sheffield, S10 2TN, UK\\
$^{5}$ Department of Astronomy \& Astrophysics, University of Toronto, 50 St. George Street, Toronto, Ontario, M5S 3H4, Canada \\
$^{6}$ Department of Physics, The University of Warwick, Coventry, West Midlands, CV4 7AL, UK\\
$^{7}$ Max Planck Institute for Gravitational Physics (Albert Einstein Institute), Hannover, Callinstra{\ss}e 38, D-30167 Hannover, Germany\\
$^{8}$ Leibniz Universit\"{a}t Hannover, 30167 Hannover, Germany\\
$^{9}$ Departamento de Astrof\'{i}sica, Universidad de La Laguna, E-38206 La Laguna, Tenerife, Spain
}
\date{Accepted XXX. Received YYY; in original form ZZZ}

\pubyear{2021}
\begin{document}
\label{firstpage}
\pagerange{\pageref{firstpage}--\pageref{lastpage}}
\maketitle

% Abstract of the paper
\begin{abstract}  
We present ULTRACAM multiband optical photometry of two transitional millisecond pulsars, PSR J1023+0038 and PSR J1227$-$4853, taken while both were in their radio pulsar states. 
The light curves show significant asymmetry about the flux maxima in all observed bands, suggesting an asymmetric source of heating in the system. We model the light curves using the Icarus binary code, using models with an additional ``hot spot'' heating contribution and an anisotropic heat redistribution model to treat the asymmetry. Our modelling reveals companion stars with under-filled Roche lobes in both PSRs J1023+0038 and J1227$-$4853, with Roche lobe filling factors in the range $f \sim 0.82-0.92$. While the volume-averaged filling factors are closer to unity, significant under-filling is unexpected from tMSPs as they must rapidly over-fill their Roche lobes to start transferring mass, which occurs on timescale of weeks or months. We discuss the motivation and validity of our extensions to the models and the implications of the under-filled Roche lobe, and suggest future work to further investigate the role of the filling factor in the tMSP cycle.
\end{abstract}

% Select between one and six entries from the list of approved keywords.
% Don't make up new ones.
\begin{keywords}
binaries:~close -- stars:~neutron -- pulsars:~individual:~PSR J1023+0038 -- pulsars:~individual:~PSR J1227$-$4853 -- stars:~evolution
\end{keywords}

\section{Introduction}
\subsection{Spiders and transitional millisecond pulsars}
Transitional millisecond pulsars (tMSPs) are a class of neutron star binary containing a recycled millisecond pulsar (MSP), spun up by accretion from a low-mass, semi-degenerate companion to spin periods of the order of milliseconds \citep{Alpar1982}. tMSPs are unique in that they are observed to transition between an accretion-powered (AP) Low-Mass X-ray Binary (LMXB) state and a rotation-powered (RP) radio pulsar state, the latter so far associated with the `redback' class of pulsars \citep{Archibald2009}. Redbacks are a sub-class of the eclipsing `spider' binaries, in which a low-mass ($0.2 {\rm M}_{\odot} \lesssim M_{c} \lesssim 0.4 {\rm M}_{\odot}$) quasi-main sequence companion star in a tight ($\sim$few hour), tidally locked orbit is irradiated by the wind of a MSP. This results in the ablation of the companion's surface into a tail of ionised matter, causing long eclipses at radio frequencies, and distinctive quasi-sinusoidal optical modulation caused by heating of the inner face of the companion, e.g. \citet{Breton2013,Roberts2011}. Spider binaries host some of the most massive and fastest spinning neutron stars \citep{Linares2020}.

As summarised in \citet{Britt2017}, observations of the three confirmed tMSP systems have revealed several shared characteristics, though it is important to note that due to the small sample size these could be coincidental. The optical emission in the RP state is indistinguishable from that of non-tMSP redback systems as described in the previous paragraph, while the AP state emission exhibits bimodal flickering and flaring \citep{Kennedy2018,Shahbaz2018b}. In the AP state, bimodal flickering and flaring is also visible in the X-ray, which indicates that an accretion disc is present \citep{Patruno2014,Linares2014,Bogdanov2015}. This is further confirmed by the strong emission lines in the optical spectrum seen in the AP state \citep{Archibald2009, Bassa2014,CotiZelati2014} which fully disappear in the RP state. In the AP state, tMSPs exhibit a flat radio spectrum suggesting self-absorbed synchrotron emission \citep{Deller2015,Bogdanov2018}, while in the RP state the radio emission is pulsed with a spectrum characteristic of synchrotron emission with a steep power law \citep{Archibald2009,Patruno2014}, typical of rotation-powered MSPs.

\subsection{The tMSP-LMXB link}
tMSPs present a unique opportunity to not only study the accretion mechanism of LMXBs, but also gain insight into the evolution of pulsar binary systems. It is generally agreed that LMXBs are the predecessor to spiders and several other types of MSP binary, but the mechanism by which the accretion is `switched off' is not known \citep{Chen2013}, nor is the mechanism by which the MSP magnetic field decays as it gets recycled \citep{Konar1997,Cumming2001}. As such, the study of tMSPs is important in uncovering the evolutionary history of spiders and LMXBs: they may be a missing link between these two populations \citep{Archibald2009,Papitto2013}. However, it is possible that they are themselves a distinct population; in this case they remain important astrophysical laboratories to study the accretion process. Since the timescale of their transitions is on the order of weeks or months, with transitions occurring every few years, we can study the entire accretion process on human timescales.

We present new optical light curves of two tMSPs, both in the radio pulsar states; PSRs J1023+0038 and J1227$-$4853. These are two of the three confirmed tMSPs; the third is PSR J$1824-2452$I, although its location in a globular cluster prevents a detailed study in optical wavelengths \citep{DeFalco2017,CotiZelati2019}. We note that there are a few `candidate' tMSPs, such as 3FGL~J$0427.9-6704$ \citep{Strader2016,Kennedy2020}, which show similar AP state properties to confirmed tMSPs but lack a radio MSP association and have not yet been seen to transition.

\subsection{PSR J1023+0038}
Often referred to as the canonical tMSP, PSR J1023+0038 (herafter J1023) was initially classified in 2001 as a cataclysmic variable system with a binary period of 0.198 days (4.75 hours) \citep{Bond2002}. The double-peaked emission lines and blue optical spectrum indicated an accreting binary with a white dwarf primary, with optical photometry showing the flickering and flaring typical of an accretion disc. \citet{Woudt2004} and \citet{Thorstensen2005} presented the first evidence for a state change, respectively showing optical photometry and spectroscopy which lacked the usual signatures of an accretion disc. The strong emission lines in the optical spectra were replaced by absorption features, while the flickering and flaring in the light curve were no longer present. The state change was confirmed in 2007 with the detection of a radio pulsar with a spin period of 1.69 ms \citep{Archibald2009, Wang2009}.

In June 2013, the radio pulsations from the MSP could no longer be observed \citep{Stappers2014}, and were replaced by rapidly varying X-ray flux \citep{Patruno2014} and optical signatures \citep{Takata2014,CotiZelati2014} indicative of an accretion disc. Kepler-K2 optical observations in 2017 further show clear evidence of an accretion disc with the slightly asymmetric, sinusoidal orbital light curve modulation still visible  \citep{Kennedy2018,Papitto2018}. J1023 displays several further characteristics not yet seen in other tMSPs.\citet{Ambrosino2017} report observations of optical pulsations of J1023 in its AP state, originating from inside the magnetosphere, and the flickering behaviour seen in the AP state is observed to occur simultaneously in optical and X-ray wavelengths \citep{Papitto2019}. Near-infrared flaring is also seen \citep{Baglio2019}. Lastly, pulsed X-ray and UV emission has been detected \citep{Jaodand2016,Jaodand2021}.

\subsection{PSR J1227$-$4853}
Identified as a variable X-ray source with XMM-Newton \citep{Bonnet2012,deMartino2012}, XSS J12270$-$4859 (now PSR J1227$-$4853, hereafter J1227) was initially classified as an LMXB due to the presence of flares and `dips' in the X-ray light curve. Between 2012 and 2013, the X-ray and optical fluxes of J1227 were observed to decrease to new minima \citep{Bogdanov2014,Bassa2014}, and the spectral emission features of an accretion disk disappeared. Radio observations revealed a MSP with a period of 1.69 ms at the source coordinates \citep{Roy2015}, showing that J1227 had transitioned from an LMXB state to radio pulsar state displaying a redback-like optical modulation with an orbital period of 0.288 days (6.91 hours). Gamma ray pulsations at the radio MSP period were discovered using data from the Large Area Telescope (LAT) on the \textit{Fermi Gamma-ray Space Telescope} (Fermi-LAT), which indicated an LMXB to a tMSP transition epoch of 2012-11-03 \citep{Johnson2015}.

\subsection{Summary of this work}
In this study we first present our new photometry, outlining the reduction and calibration procedure, in section \ref{sec:opticalobs}. In section \ref{sec:asymm} we discuss the nature of and the potential mechanisms behind the asymmetry of the light curves. We discuss our modelling of these light curves using the \texttt{Icarus} binary light curve synthesis code in section \ref{sec:numericalmodelling}, in particular our constraints on the orbital parameters of the systems, and implement two extensions to the \texttt{Icarus} model. The first extension accounts for an additional hot spot on the surface of the companion, and the second is a new description of the temperature distribution of the companion which takes into account diffusion and convection in the outer shell. Then, we outline the results of the modelling and discuss their validity and the implications on the tMSP transition mechanism in sections \ref{sec:results} and \ref{sec:discussion}.

\section{Optical Observations} \label{sec:opticalobs}
\subsection{ULTRACAM on the NTT}
\label{sec:UCAMNTT}
Our observations were performed using the ULTRACAM instrument mounted on the 3.5 m New Technology Telescope (NTT) at the La Silla observatory, Chile. ULTRACAM \citep{Dhilon2007} is an optical imaging photometer capable of simultaneous 3-band observation. We used filters from the ULTRACAM Super Sloan Digital Sky Survey (Super-SDSS) u$_{\rm s}$g$_{\rm s}$r$_{\rm s}$i$_{\rm s}$z$_{\rm s}$ photometric system \citep{Dhillon2018}, with u$_{\rm s}$ and g$_{\rm s}$ filters on the first two CCDs, and either of i$_{\rm s}$ or z$_{\rm s}$ for the third. Our typical integration time was 10 seconds with 25 ms dead time between each frame. Our observations are summarised in table \ref{table:observations}.

\begin{figure}
    \centering
    \includegraphics[width=0.49\textwidth]{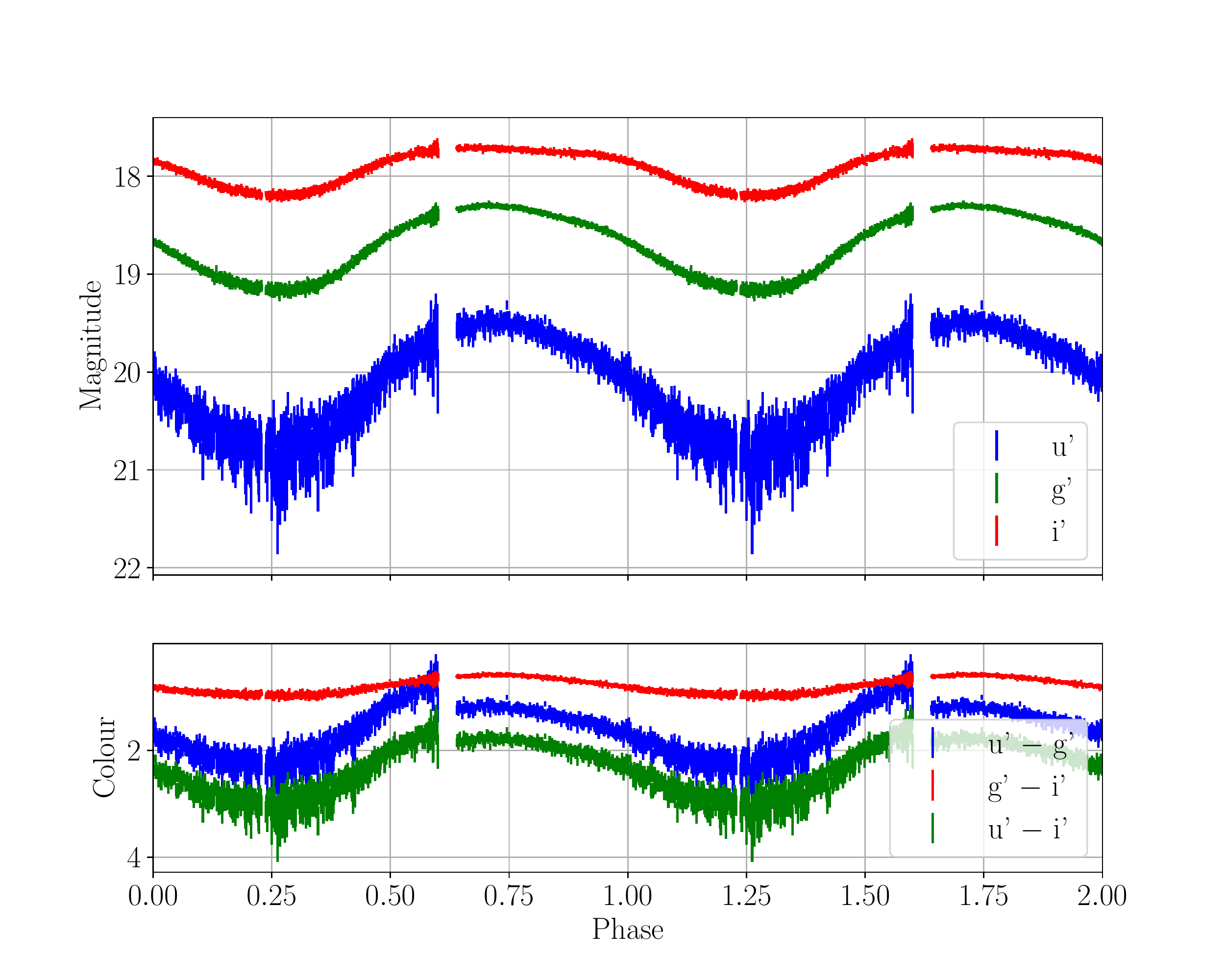}
    \caption{Phased light curve of J1227, repeated over two cycles for clarity, with each colour corresponding to a different filter as in the legend. The asymmetry can clearly be seen in the i$_{\rm s}$\ and g$_{\rm s}$\ bands around phase 0.75. At phase 0.6, the Sun rising is responsible for the large scatter, particularly so in the u$_{\rm s}$\ band. The bottom panel shows the colour information. We adopt the phase convention where the pulsar is at superior conjunction at phase 0.25.}
    \label{fig:J1227_lc}
\end{figure}

\begin{figure}
    \centering
	\includegraphics[width=0.49\textwidth]{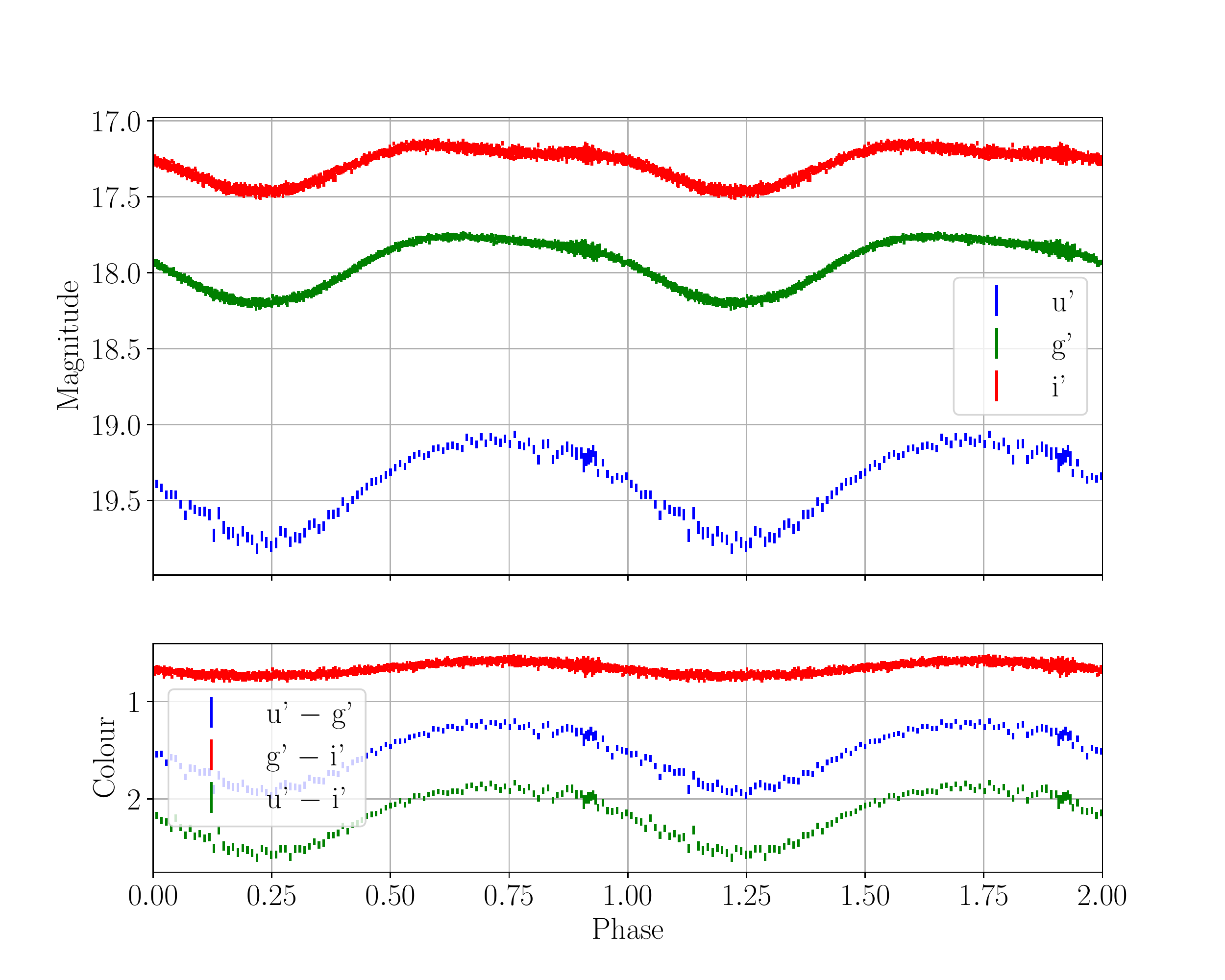}
    \caption{Phased light curve of J1023, clearly showing the asymmetry around phase 0.75. Similarly to J1227, this asymmetry is more pronounced in the i$_{\rm s}$ and g$_{\rm s}$ bands. Note that the artefacts around phase 0.8-0.9 are due to poor seeing conditions.}
    \label{fig:J1023_lc}
\end{figure}

\begin{center}
\begin{table*}
 \begin{tabular}{c  c  c  c  c  c  c  c  c} 
 \hline
Date & \multicolumn{1}{p{1.4cm}}{\centering Start time \\ (UTC)} & Source & Exposures & \multicolumn{1}{p{1.3cm}}{\centering Phase \\ coverage} & Seeing (") & Airmass & u$_{\rm s}$ co-adds & \multicolumn{1}{p{1.3cm}}{\centering Exposure \\ time (s)}\\ [1ex]
 %Model Parameters & $i=50$\textdegree & $i=60$\textdegree & $i=70$\textdegree \\ [0.5ex] 
 \hline\hline
 \\[-1em]
 2010-05-04 & 23:03:48 & J1023 & 1859 / 1858 / 102 & $\sim105$\% & 1.23 & 1.15-2.08 & 18 & 9.56 \\
 %\\[-1em]
  %2010-05-05 & 22:58:12 & J1023  & u$_{\rm s}$ / g$_{\rm s}$ / i$_{\rm s}$ & $\sim 90$\% & 1.14 & 1.15-2.16 & 6 & 9.56 \\
 %\\[-1em]
  %2010-05-06 & 23:14:26 & J1023  & u$_{\rm s}$ / g$_{\rm s}$ / z$_{\rm s}$ &  $\sim105$\% & 1.17 & 1.15-2.26 & 6 & 5.76 \\
 %\\[-1em]
  2019-02-27 & 03:04:41 & J1227  & 2360 / 2359 / 1180 & $\sim90$\% & 0.90 & 1.06-1.42 & 2 & 10.02 \\
% \\[-1em]
  %2020-01-27 & 04:59:54 & J1227  & u$_{\rm s}$ / g$_{\rm s}$ / i$_{\rm s}$ & $\sim 60$\%& 1.14 & 1.06-1.45 & 3 & 10.01 \\
 %\\[-1em]
  %2020-01-28 & 05:10:59 & J1227  & u$_{\rm s}$ / g$_{\rm s}$ / i$_{\rm s}$ & $\sim 55$\% & 1.04 & 1.06-1.39 & 3 & 10.01 \\
 %\\[-1em]
  \\[-1em]
 \hline
\end{tabular}
\caption{Table of observations of J1227 and J1023, with information gathered from the ULTRACAM online logs and reduction files. The seeing, displayed as the median, was calculated from the FWHM of the observations. Both nights used the u$_{\rm s}$ / g$_{\rm s}$ / i$_{\rm s}$ configuration of filters; the number of exposures in the table correspond to these filters.  Further to the co-adds shown here, no further binning in time was performed during reduction.}
 \label{table:observations}
\end{table*}
\end{center}

\subsection{Observations of J1227}
\label{sec:J1227obs}
J1227 was observed on 2019-02-27 beginning at 03:09:53 UTC, during its radio pulsar state. The observations were completed in one night, providing more than 90\% orbital phase coverage in mostly photometric conditions, although some clouds were present near the end of the observation, decreasing the SNR of these images. We reduced the data with the ULTRACAM pipeline using an ensemble aperture photometry method \citep{HoneyCutt1992}. We used 8 comparison stars common to u$_{\rm s}$, g$_{\rm s}$, and i$_{\rm s}$ to correct for atmospheric transmission variations. We employed the same 8 calibration stars of known i$_{\rm s}$\ and g$_{\rm s}$\ magnitudes from the AAVSO Photometric All-Sky Survey (APASS) to calibrate the i$_{\rm s}$\ and g$_{\rm s}$\ magnitudes to the absolute photometry system\footnote{\href{https://www.aavso.org/apass}{https://www.aavso.org/apass}}. The same comparative photometry was also performed for the u$_{\rm s}$\ band, but as there were no objects with known u$_{\rm s}$\ magnitude in the field we used the zero point of this band, calculated from separate observations of SDSS standard stars, to calibrate the magnitudes instead. While the zero point and typical extinction coefficients are known for ULTRACAM in this configuration, this is a less accurate calibration method than comparative photometry, so we included a larger band calibration offset for the u' band in the modelling. 

During observations a temporal co-addition factor of 2, where the CCD is read out every other exposure, was used for the u$_{\rm s}$\ band and the resulting SNR of the data was sufficient that we did not need to perform any further temporal averaging of any of the bands. The final step of our reduction was to discard any observations with error flags from the pipeline or SNR below a threshold of 3. This second condition was used as several observations near the optical minimum were impacted by cloud cover. The resulting dataset contains a total of 5899 good data points: 2360 in i$_{\rm s}$, 2359 in g$_{\rm s}$, and 1180 in u$_{\rm s}$. These data were folded at the orbital period using the ephemerides from radio timing \citep{Roy2015}. We will apply the following convention thorough the paper to define the orbital phase: phase 0.0 corresponds to the pulsar ascending node (i.e. for a circular orbit this is defined as the quadrature point when the pulsar is moving away from us), with the companion's inferior conjunction (optical minimum) occurring at phase 0.25.

The phased light curve of J1227 (see fig. \ref{fig:J1227_lc}) displays single-peaked sinusoidal modulation, due to the irradiation of the companion. The peak-to-peak amplitude of modulation is approximately 0.6 mag in i$_{\rm s}$, 0.8 mag in g$_{\rm s}$, and 1.4 mag in u$_{\rm s}$, with mean magnitudes of 18.0 mag, 18.7 mag, and 20.4 mag respectively. Considering the colour information, the companion star becomes redder during the pulsar superior conjunction (optical minimum), in line with the expectation that the night side of the star is cooler than the day side. The light curve shows the asymmetric nature of the modulation, and a `flattening' of the optical maxima most noticeable in the i' band due to a significant ellipsoidal modulation contribution.

\subsection{Observations of J1023}
\label{sec:J1023obs}
J1023 was observed over 3 consecutive nights starting on 2010-05-04, during the object's radio pulsar state. While the observations provide nearly 100~per~cent phase coverage in u$_{\rm s}$, g$_{\rm s}$, i$_{\rm s}$, and z$_{\rm s}$, the latter two nights suffered from cloudy skies and so the quality of the first night of observations far exceeds that of the second and third, both in terms of usable phase coverage and SNR. Additionally, due to the cloud cover, the magnitude calibration is not completely consistent between nights. As such, we performed the modelling using just the data from 2010-05-04 to ensure that these potential calibration issues did not affect the results.

The observations were reduced in the same way as with J1227, using ensemble aperture photometry with the ULTRACAM pipeline. However, as the position of J1023 has been covered by the Sloan Digital Sky Survey (SDSS), calibration stars were available for all four bands, including u'. In total, 12 comparison stars were used for i$_{\rm s}$\ and z$_{\rm s}$, 11 for g$_{\rm s}$, and 6 for u$_{\rm s}$, with the same number being used to calibrate the magnitudes. We obtained a total of 3819 good data points; 1859 in i$_{\rm s}$, 1858 in g$_{\rm s}$, and 102 in u$_{\rm s}$, which were folded on the orbital period.

The phased light curve, shown in figure \ref{fig:J1023_lc} shows asymmetrical modulation in all 3 bands, with a single irradiation peak in u$_{\rm s}$. The relative contribution of the ellipsoidal variation to the light curve shape, which produces the double-peaked modulation per orbit (see e.g. \citet{Li2014} for a clear example of this), is higher for redder bands (see e.g. i$_{\rm s}$ band compared to u$_{\rm s}$ band). We measure mean magnitudes of 17.3 in i$_{\rm s}$, 17.9 in g$_{\rm s}$, and 19.4 in u$_{\rm s}$, with modulation amplitudes of 0.3 mag, 0.4 mag, and 0.7 mag respectively.

\subsection{Radial Velocities}
\label{sec:radvel}
To help constrain the projected companion radial velocity, $K_2$, of J1023 we combined our photometric data with spectroscopic radial velocity measurements from \citet{Shahbaz2019} and \citet{McConnell2018}. We used radial velocity curves obtained from metallic line spectra captured with the ISIS instrument on the 4.2 m William Herschel Telescope (WHT) in 2016 for \citet{Shahbaz2019} and 2009 for \citet{McConnell2018}. Both radial velocity curves had been produced using broadly the same set of metallic lines, over the same range of wavelengths.

\begin{figure*}
	\includegraphics[width=0.85\textwidth]{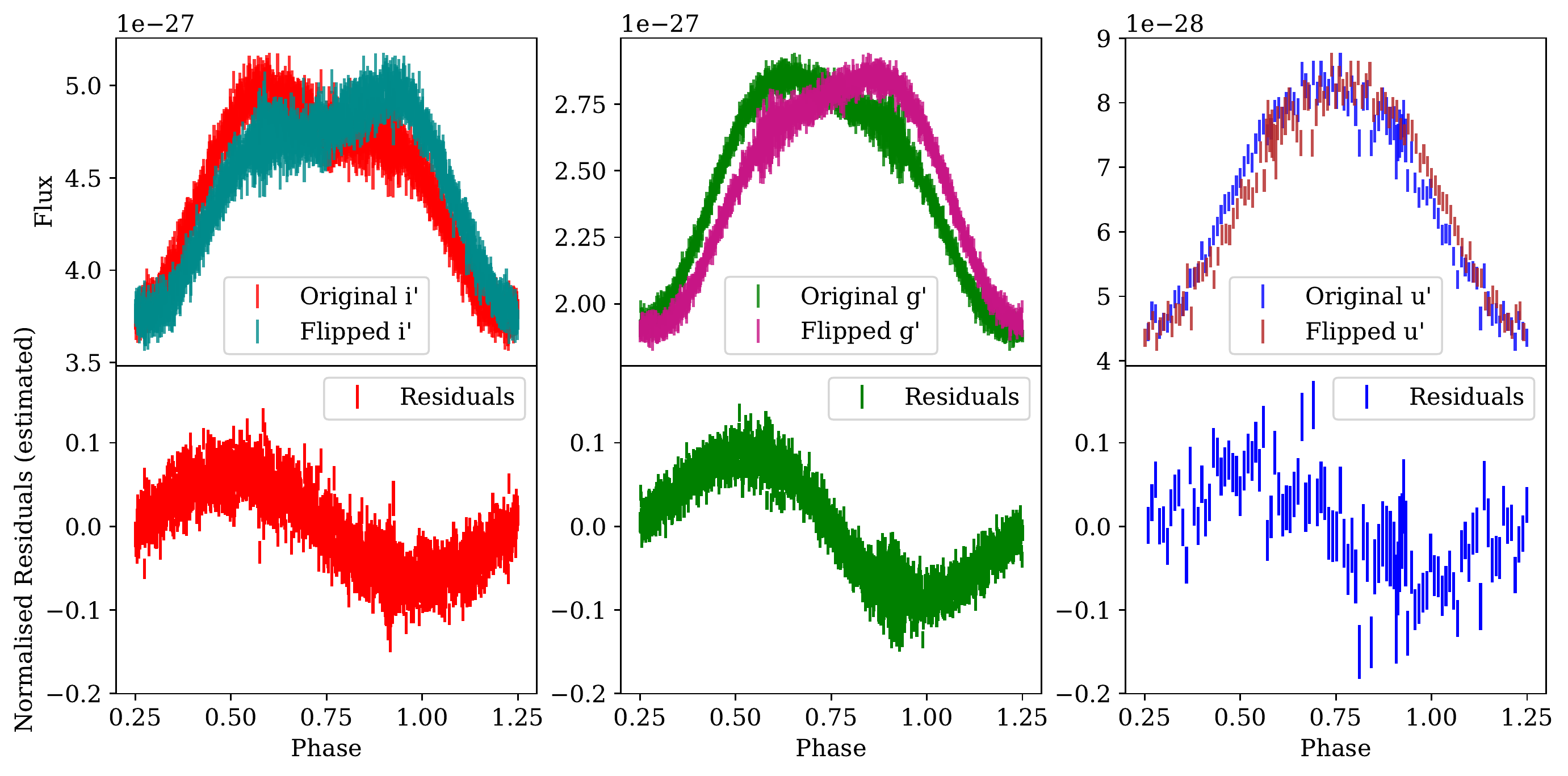}
    \caption{\textit{Top: } i$_{\rm s}$, g$_{\rm s}$, and u$_{\rm s}$\ light curves of J1023 overlaid with the same light curves mirrored about phase 0.5, illustrating the asymmetries. \textit{Bottom: } Estimated residuals between original and mirrored light curves, normalised to the mean band flux. The mirrored light curve was interpolated over the original phases in order to calculate these residuals.
    }
    \label{fig:symmetriesJ1023}
\end{figure*}

\section{Asymmetries}
\label{sec:asymm}
While observed in both these tMSP systems, asymmetric light curves appear to be a feature of redback systems in general (e.g. PSR J2215+5135, \citet{Schroeder2014}) and are not specific to tMSPs. They are also not unique to either the rotation powered state or accretion powered state of a tMSP, as the asymmetry visible in the optical light curve of J1023 presented in this work during the rotation powered state is also visible during the accretion powered state (\citealt{Kennedy2018}; \citealt{Papitto2018}). As such it is unlikely that they arise from, for example, reprocessing or obscuring of the pulsar wind by some disc remnant. Indeed there is no mechanism driving the asymmetry that is widely accepted and evidenced, though there are a number of possible theories.

Considering the work of \citet{Romani2016}, a swept-back intra-binary shock (IBS) between the pulsar and companion winds could be responsible for the asymmetry via heating of the companion by non-thermal X-ray emission produced in the wind shock. In their work, the modelling includes the effect of the IBS heating on the companion and finds good agreement with data. More recently, \citet{Kandel2020} performed modelling of the asymmetric redback PSR J2339-0533 using a hot spot model which aims to describe the ducting of high-energy particles (such as those shed from the IBS) onto magnetic caps on the companion star's surface.

Despite the relative success of the models linked to the IBS, the results of \citet{Zilles2020}, which estimate the penetration depths of high energy photons in the companion photosphere, suggest that the X-rays reprocessed by the shock could not sufficiently heat the companion to the observed asymmetry temperatures.

Dynamics on the companion surface may instead produce an asymmetric temperature profile. As the day side of the companion is strongly heated, we may expect strong circulatory winds and thermal structures similar to those observed on hot Jupiters (see, e.g., \citet{Komacek2020,Jackson2019}) or cataclysmic variables \citep{Martin1995}. The lack of fusion in hot Jupiters allows these global winds to form complex meteorologies, which are unlikely to form on redback companions as the radial convection in the envelope would disrupt it. The large temperature gradient between the day and night side would be sufficient to fuel the winds and so allow heat to flow through the atmosphere, with circulation driven by the spin of the companion. We implement this as presented in \citet{Voisin2020}, however we note there are possible alternatives. In \citet{deWit2012}, an alternative hot spot model is presented, with the temperature distribution motivated by these thermal flows. Or, in \citet{Demory2013}, a longitudinal temperature map of the surface of the companion is used, using a number of fixed bands.

A recent, novel approach in \citet{Romani2021} models the asymmetry of the optical light curve of the black widow pulsar PSR J1810+1744 by acting on the gravity darkening parameter. However, this method was used in the case of a very highly irradiated companion, while these tMSPs display only modest irradiation compared to their internal energy output.

We note the similarity to the HW Vir class of compact binaries, consisting of a hot sub-dwarf primary and cool, close companion (typically a white dwarf; \citet{Schaffenroth2019}), which do not display asymmetric light curves despite the otherwise similar orbital parameters. In these systems the sub-stellar point on the companion is heated to temperatures of over $\sim~10^4$~K, significantly more than the tMSPs in this work. We suggest that the depth to which the irradiation penetrates the companion's photosphere in HW Vir binaries is significantly shallower than with tMSPs, due to the different source of heating. As such  the heat redistribution layer may not be sufficiently deep to produce asymmetries.

To illustrate the asymmetry in J1023 we overlay the light curves of each band, in fluxes, over the same light curve mirrored about phase 0.5. We are then able to analyse the asymmetry by calculating the residuals. We normalise these residuals to the mean flux of each band then interpolate the flipped light curve at the phases of the original in order to calculate residuals between each curve. Note that this means that these are not residuals in the traditional sense due the interpolation, however they clearly demonstrate the difference between the original and mirrored light curves. Seen in figure \ref{fig:symmetriesJ1023}, the residuals follow a nearly sinusoidal shape across all three bands. The amplitude of these sinusoids is also comparable across the bands, largest in the g$_{\rm s}$\ band, followed closely by i$_{\rm s}$\ and then u$_{\rm s}$\ at around 60~per~cent the amplitude of i$_{\rm s}$. Note that there is significantly increased scatter in u$_{\rm s}$\ compared to g$_{\rm s}$\ and i$_{\rm s}$, which causes the amplitude of the sinusoid to appear larger than it is. 

\begin{figure*}
    \centering
    \includegraphics[width=0.85\textwidth]{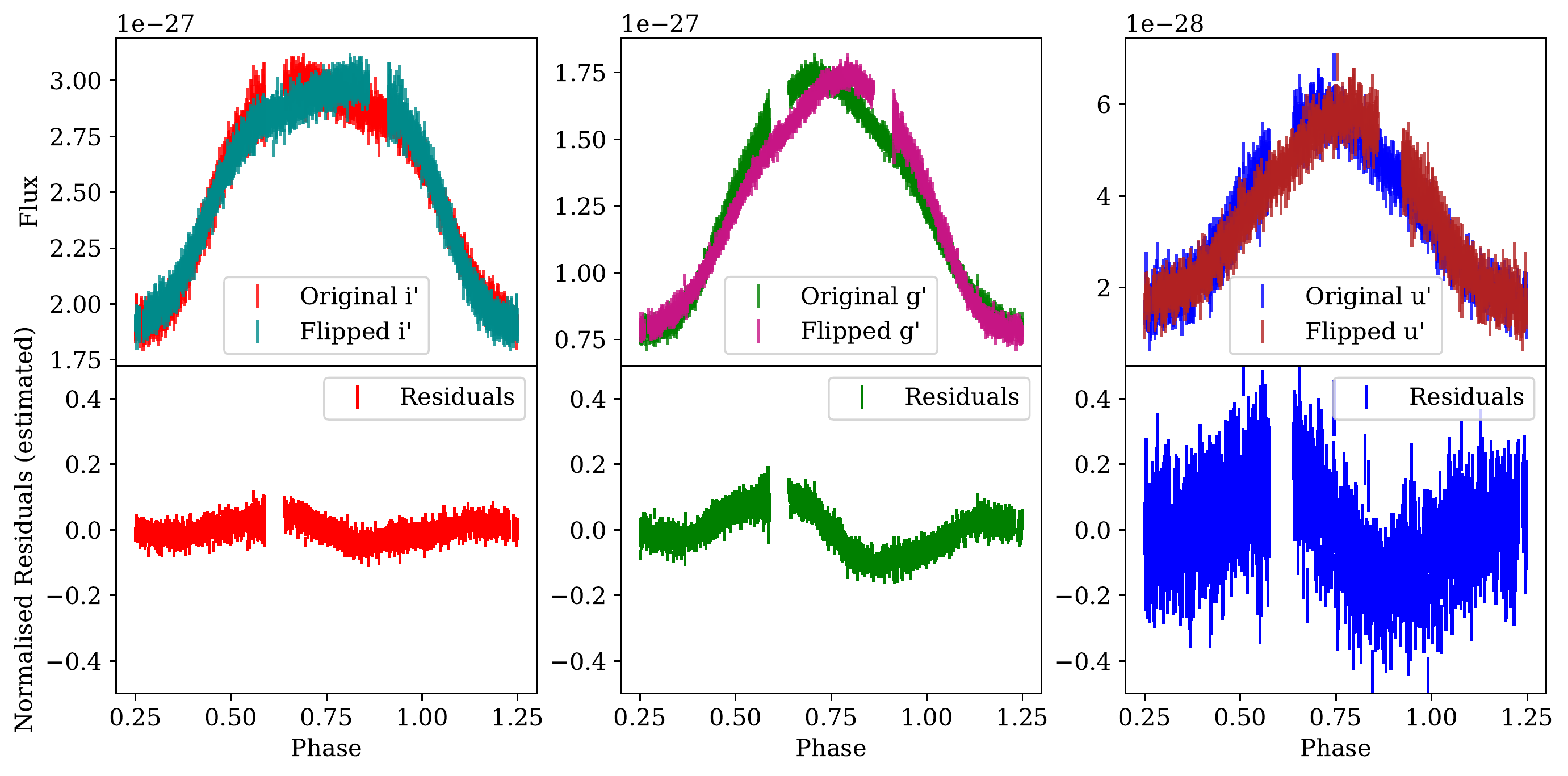}
    \caption{As figure \ref{fig:symmetriesJ1023}, for J1227. The light curves were trimmed to remove the region of large scatter around phase 0.6 to. Note the shape now reflects that of a second-harmonic sinusoid, compared to the fundamental sinusoid seen in J1023.}
    \label{fig:symmetriesJ1227}
\end{figure*}

Performing the same analysis on the light curves of J1227, we find that the shape of the residuals is again consistent between each band. However, they more closely follow the shape of a sinusoid at the second harmonic; this is shown in figure \ref{fig:symmetriesJ1227} for the g$_{\rm s}$\ band. Additionally, the amplitude of this modulation is much more varied between bands; strongest in u$_{\rm s}$, decreasing to 60~per~cent in g$_{\rm s}$ and finally to roughly 25~per~cent in i$_{\rm s}$. 

\section{Numerical Modelling}
\label{sec:numericalmodelling}
\subsection{Icarus}
\label{sec:icarus}
We modelled our optical light curves using the Icarus code (see \citealt{Breton2012} for a thorough introduction) in order to constrain the orbital parameters of the system and the temperature profile of the irradiated companion. We use atmosphere grids created using the ATLAS9 synthesis code \citep{Castelli2003}. In the standard heating model, the input parameters the model uses are as follows: the orbital inclination angle, $\cos(i)$, the Roche lobe (RL) filling factor, $f$, the base (night side) temperature of the companion, $T_0$, the irradiation temperature of the companion, $T_{\rm irr}$, the distance modulus, DM, the companion's projected radial velocity amplitude, $K_2$, the mass ratio (defined as the ratio of the pulsar mass to the companion mass), $q$, the co-rotation coefficient, $\Omega$, the gravity darkening coefficient, $\beta_{\rm g}$, and the V-band extinction coefficient, $A_v$. Note that while the Icarus code uses the DM as a model parameter, in this work we discuss the distance which is derived from this.

The filling factor is defined as the ratio of the stellar surface radius in the direction of the pulsar to the distance to the L1 point. We also derive a volume-average filling factor which is a representation of the volume of the star to the volume of the Roche lobe, $f_{\rm VA} = \langle R \rangle / \langle R_{\rm RL} \rangle$. We fix the co-rotation coefficient to $\Omega = 1$, as we assume both systems are tidally locked, and for both sources we fix $\beta_{\rm g} = 0.08$ as we assume the late-type companion stars have large convective envelopes \citep{Lucy1967}. As well as the volume-averaged filling factor and distance, we also derive the pulsar mass, $M_{\rm psr}$ and the blackbody-equivalent temperatures of the day and night side of the companion.

Relations connecting some of the parameters previously introduced and the orbital ephemerides are also considered in our modelling. In particular, the mass ratio follows the relation
\begin{equation}
    \label{eq:qFromK}
    \centering
    q = \frac{K_2 P_B}{2 \pi a_1},
\end{equation}
where $P_B$ is the orbital period and $a_1$ is the projected semi-major axis of the pulsar, derived from radio timing (\citet{Roy2015} and \citet{Archibald2009} for J1227 and J1023 respectively).

Finally, we allow the systemic velocity, $\Gamma_v$ to be a free parameter as our radial velocity data are not mean-subtracted. At each step in the MCMC chain, we use the Icarus code to determine the effective centre-of-light radial velocity, $v_{\rm eff}$, of the companion, evaluated using a model atmosphere corresponding to the wavelength range of the spectroscopic radial velocity curve. In this case, this corresponds to the SDSS r' band. The amplitude of this effective radial velocity curve is defined as $K_{\rm eff}$. We then fit for and subtract any linear offset between the modelled and measured radial velocity curves, and calculate the residuals. These residuals are combined with the residuals from the modelled light curve to calculate the posterior probability at each step in the chain.

To determine the model parameters we used the multi-nested sampler, \texttt{MultiNest} \citep{Feroz2009}, implemented in Python as \texttt{pymultinest} \citep{Buchner2014}. This method also allows us to directly compare the Bayesian evidence of each choice of model. In this paper this is quoted as the natural logarithm of the model evidence, $\log Z$, and the reduced $\chi^2$ is determined from the best posterior solution (as opposed to the best likelihood solution).

The selection of priors is very important as the model is extremely degenerate; where possible, we use priors on parameters informed by published measurements. For these we use Gaussian priors, centred on the literature value with standard deviation equal to the given uncertainties (i.e. the 68~per~cent significance). For parameters with no known constraints, we either use top-hat priors over a range of physically sensible values (such as constraining the temperature to the range of the atmosphere grids) or leave the parameter unconstrained.

\subsection{Standard symmetrical direct heating model}
\label{sec:symmmodel}
Initially, we used a symmetrical direct heating model to act as a benchmark. This model assumes a constant base temperature, $T_0$, across the companion, then models the effect of heating by assuming that the additional flux is thermalised and locally re-emitted such that we can express the day side temperature, $T_{\rm day}$ as
\begin{equation}
    T_{\rm day}^4 = T_0^4 + T_{\rm irr}^4,
\end{equation}
where $T_{\rm irr}$ is the so-called irradiation temperature. The effect of gravity darkening is applied prior to the irradiation to give the temperature distribution of the companion. Irradiation effects take into account the distance between the companion and the pulsar and the incidence angle of the irradiation.

For J1227 the free parameters in this initial model were the orbital inclination, $\cos i$, the base temperature, $T_0$, and irradiation temperature, $T_{\rm irr}$, of the companion, the distance modulus, $DM$, the companion velocity $K_2$, and the filling factor, $f$. While \citet{deMartino2014} suggest an inclination of between $43\text{\textdegree} < i < 73$\textdegree\ and the modelling of \citet{deMartino2015} constrains $46\text{\textdegree} < i < 65$\textdegree, we opted to leave the inclination unconstrained to perform independent modelling (i.e. not biased by previous studies).

We used top hat priors on the temperatures, setting the limits in accordance with the range of temperatures covered by our model atmosphere grids; 1300 K to 10000 K. We left the filling factor mostly unconstrained, with limits $0.0 < f < 1.0$, as we expect the Roche Lobe to be mostly full, but have no physically imposed minimum filling factor.

The priors on the distance modulus were calculated following the method described in \citet{Luri2018}, using the joint probability distribution of distances derived from the GAIA parallax measurement combined with the model of galactic MSP densities from \citet{Leavin2013}. The GAIA parallax was significant, at $0.623 \pm 0.168$ mas, and so this dominated the distribution.

We constrain the companion velocity, $K_2$, based on the radial velocity amplitude inferred from spectroscopy by \citet{deMartino2014}; $K_2 = 261 \pm 5$ km s$^{-1}$. Note that the radial velocity data are not publicly available at the time of writing. Rather than using a simple Gaussian prior on this value, we instead use the radial velocity curve method described in the previous section. Our radial velocity `curve' consists of this single value at phase $\phi = 0.25$, the pulsar inferior conjunction, where the maximum projected velocity occurs. This method ensures the radial velocity amplitude from spectroscopy is corrected to the centre of mass of the companion.

We also use Gaussian priors on the derived parameter $V\sin(i)~=~86~\pm20$~km~s$^{-1}$, the companion rotational velocity, obtained from the same spectroscopy, using the relation derived in \cite{Wade1988},
\begin{equation}
    V\sin(i) = (K_1 + K_2) R_2(f),
\end{equation} 
where $K_1$ is the radial velocity of the pulsar, and $R_2(f)$ is the volume-averaged radius of the companion star (in units of $a$) evaluated at a given RL filling factor, $f$. We therefore use this to constrain both the filling factor and companion system velocity. We obtain $K_1$ from radio timing \citep{Roy2015}, while $K_2$ and $R_2(f)$ will be calculated by the Icarus model.

For J1023, we used the same top hat priors on $T_0$, $T_{\rm irr}$, and $f$ as with J1227. While a well-constrained inclination can be derived from the results of \citet{Deller2012}, $42 \pm 2$\textdegree, these calculations assume that the companion is Roche-lobe filling \citep{Thorstensen2005}. As such we do not use any priors on the inclination.

\citet{Deller2012} accurately determined the distance to the system from parallax measurements using long baseline radio interferometry to be $d=1.368\sus{0.042}{0.039}$~kpc. This is much more precise, with a lower uncertainty, than estimating the distance using the GAIA parallax method and so this was used to inform our distance modulus priors.

\begin{figure}
\centering
\includegraphics[width=0.99\columnwidth]{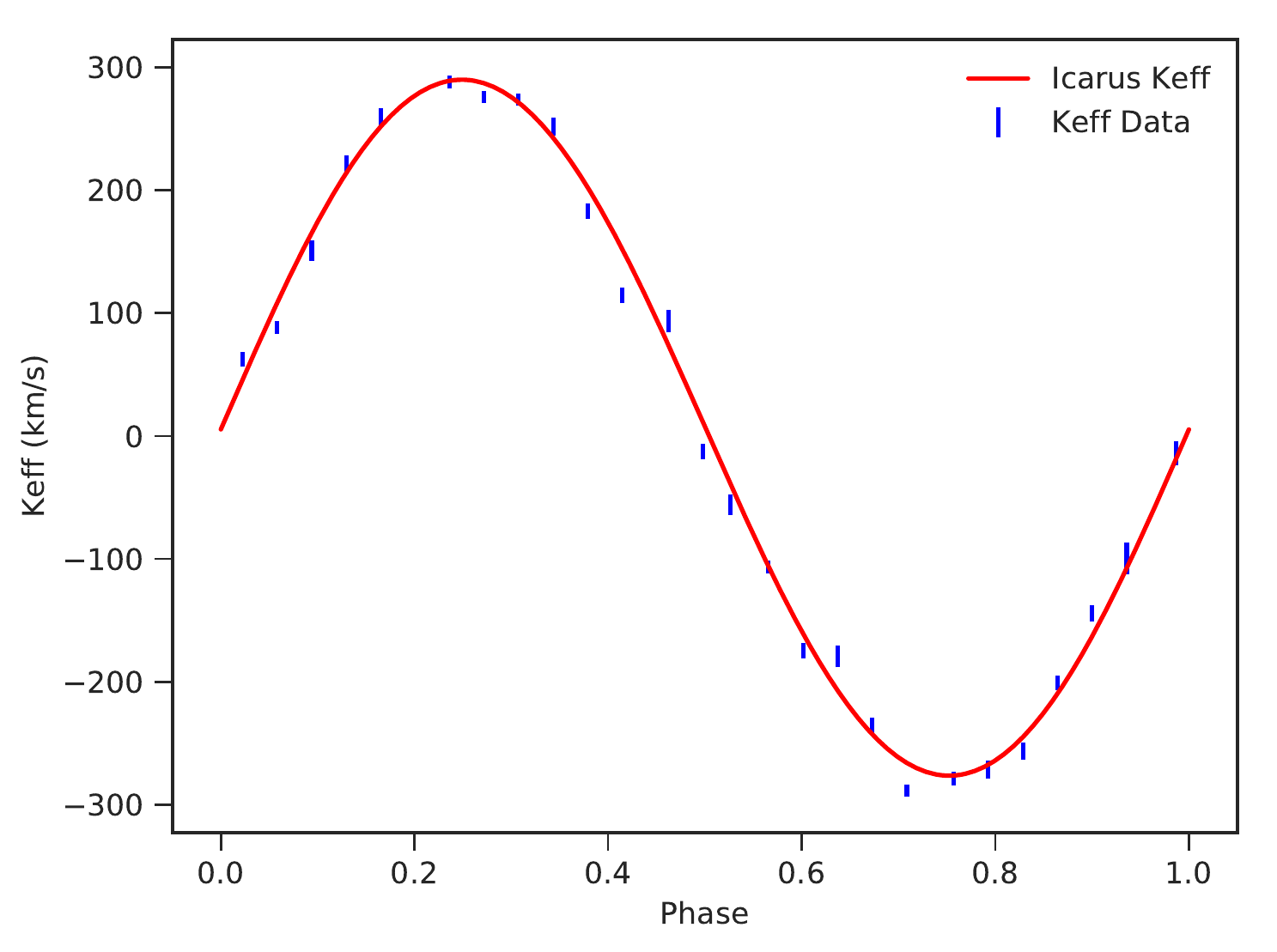}
\caption{Radial velocity fitting curve for J1023. The points in blue are the radial velocity measurements from spectroscopy \citep{Shahbaz2019}, and in red is the best-fitting radial velocity curve from Icarus.  The red RV curve is calculated at each step in the MultiNest sampler and the $\chi^2$ value from the fit to the data is added to the posterior distribution. $K_{\rm eff}$ is the effective centre-of-light radial velocity of the companion.}
\label{fig:mcmc18-keff}
\end{figure}

For $K_2$, there are radial velocity measurements available from \citet{Shahbaz2019}, however the heating of the companion distorts the radial velocity curve. This results in variable measurements of the radial velocity semi-amplitude. We take the measurements obtained from 2009 ISIS observations in the pulsar state of metallic absorption lines which correspond to a radial velocity semi-amplitude of $K_2=276.3\pm~5.6$~km~s$^{-1}$. Instead of using these to inform a Gaussian prior on $K_2$, we use the data presented in their work to fit a radial velocity curve using the method described in section \ref{sec:icarus}. Figure \ref{fig:mcmc18-keff} shows these data and the fitted radial velocity curve. From the same work we used the $V\sin(i)$ measurement of $V\sin(i)~=~77.7\pm 2.7$~km~s$^{-1}$ to constrain $f$ and $K_2$, also from the 2009 ISIS observations of metallic absorption lines. 

We also fit the V-band interstellar extinction separately for each source. For J1227 we use the prior value $A_V = 0.341$, calculated using the relationship $A_V = 3.1 E(B-V)$ \citep{Cardelli1989} from the colour excess presented in \citep{deMartino2014}, in turn calculated from the $N_H$ column density presented in  \citet{deMartino2010}. We allow for a $20$~per~cent uncertainty on this value. For J1023 we test two methods. First, we used the colour excess determined in \citet{Shahbaz2015} (again using the $N_H$ column density) to calculate the extinction and the same relationship, obtaining $A_V = 0.2263$ and allowing for the same 20~per~cent uncertainty. Second, we use the Pan-STARRS dust maps of \citet{Green2018} to obtain $A_V = 0.109$  As these methods do not produce consistent extinction coefficients, we compared the evidence of models with each $A_V$ prior and otherwise identical parameters. For both sources we use the reddening coefficients of \citet{Schlafly2011} to calculate the appropriate extinction for each band.

\subsection{Single-spot heating model}
\label{sec:spotmodel}
We extended the symmetrical model to include a single hot spot on the companion's surface. This was motivated by the successful modelling of similar asymmetries in other redback sources using a single-spot model (e.g., \citet{Shahbaz2017,Romani2016,Nieder2020,Clark2021}). Our motivation for the inclusion of a hot spot is largely empirical, however we discuss two possible physical origins of the asymmetric heating in section \ref{sec:asymm}; heating from X-rays reprocessed by a swept-back shock, and thermal winds on the companion surface.

Each spot introduces four free parameters: the spot temperature, $T_{\rm spot}$, the spot radius, $R_{\rm spot}$, and the spot position angles, $\theta_s, \phi_s$. A full description of the spot geometry and heating contribution is presented in \citet{Clark2021}, and a summary of the model extensions used in this work is included in appendix \ref{appendix:models}. We use a single spot as using more than one is likely to over-fit the data, while also being more computationally expensive and significantly increasing the degeneracy of the parameter space. The priors on the spot temperature and radius are uniform over the ranges $0\ {\rm K} < T_{\rm spot} < 10000$~K and 0\textdegree\ $<R_{\rm spot} < 90$\textdegree\ respectively. This upper bound on the spot temperature also corresponds with the maximum temperature covered by the atmosphere grids used.

We note that the asymmetric heating caused by a hot spot on the leading edge of the companion (the side of the companion moving `forwards' through the orbit) can also be described by a `cold' spot on the trailing edge (the side moving `backwards'). As such, we also performed fitting with a cold spot with a negative spot temperature, $T_{\rm spot} < 0$. We use uniform priors on $\cos \theta_s$ and uniform priors on the $\phi_s$ angle. To constrain the spot temperature and radius, we use Gaussian priors on the intensity of the flux from the spot, $I \propto T^4 R^2$, with a mean of $I=0$ and a width of $\sigma_I = 10^{12}$. This was chosen in order to avoid a very small ($R_{\rm spot} \leq 5 $ deg), very hot spot, since this prior favours cooler, larger spots.

\subsection{Heat Redistribution}
\label{sec:hr}
As detailed in section \ref{sec:results}, our modelling with both the symmetric and hot spot models produced results that were not reliable, and indeed indicated that neither model sufficiently describes the highly asymmetric light curves of these systems. As a result, we additionally used a further extension of the \texttt{Icarus} code which directly models heat redistribution via diffusion and convection within the outer envelope of the companion. We implement the treatment in \cite{Voisin2020}, though we acknowledge also the treatment of wind circulation in \citet{KandelRomani2020} as a specific case of the latter. We continue our use of MultiNest evidence sampling to constrain the model parameters and heat redistribution laws. While the direct heating model assumed a constant companion base temperature (save for the effects of gravity darkening) and the hot spot extension assumes an additional, fixed, temperature source, the heat redistribution model allows for parallel (that is, with no radial component) energy transport within the outer shell. The details of this model are presented in appendices \ref{appendix:models} and \ref{appendix:hrap}, while a full treatment can be found in \citet{Voisin2020}.

This model allows for a choice of convection profile, $f(\theta)$. We initially chose a linear profile, $f(\theta) = v$, where $v$ is the strength of the convection current in energy flux per unit temperature and is the first additional model parameter. This profile describes constant longitudinal advection. We allow for the diffusion coefficient, defined in appendix \ref{appendix:models}, to depend on the local temperature following a power law of index $\Gamma$. We performed fits with both $\Gamma = 0$ for linear diffusion, and with the diffusion index as a free parameter. The diffusion coefficient, $\kappa$, is the last additional model parameter. When $\kappa = 0$, the diffusion is switched off and the model becomes convection only. In this case, this model is equivalent to that in \citet{KandelRomani2020} when used with a `bizone' convection profile. However, we do not use that profile in this work. We performed fits with both $\kappa = 0$ and $\kappa$ as a free parameter. 

\begin{center}
\begin{table*}
 \begin{tabular}{l  c  c  c c} 
 \hline
 Parameters & Symmetrical & Single spot & HR 1 & HR2\\ [0.5ex] 
 \hline\hline
  \\[-1em]
 Inclination, $i$ (\textdegree) & $88.3\substack{+1.3 \\ -2.0}$ & $76.6\substack{+0.6 \\ -1.2}$ & $77.1 \substack{+ 0.1\\ -0.15 }$ & $77.0 \substack{+ 0.2\\ -0.3 }$ \\ 
 \\[-1em]
 Mass ratio, $q$ & $5.6\substack{+0.1 \\ -0.1}$ & $5.8\substack{+0.1 \\ -0.1}$ & $5.48\substack{+ 0.07\\ -0.07 }$ & $5.5\substack{+ 0.1\\ -0.1 }$\\
 [-1em]\\
Radial velocity, $K_{2}$ (km s$^{-1}$) & $282\substack{+5 \\ -5}$ & $294 \substack{+4 \\ -5 }$ & $277\sus{3}{4}$ & $277\sus{5}{5}$\\
 \\[-1em]
 Filling factor, $f$ & $0.825\substack{+0.002 \\ -0.002}$  & $0.838\substack{+0.003 \\ -0.003}$ & $0.850\sus{0.002}{ 0.002}$ & $0.852\sus{0.002}{ 0.002}$ \\
 \\[-1em]
 Base temp., $T_0$ (K) & $5452\substack{+19 \\ - 20}$ & $5556\substack{+14 \\ -15}$ & $5584\sus{11}{11}$ & $5585\sus{15}{15}$\\
 \\[-1em]
 Irradiation temp., $T_{\rm irr}$ (K) & $5230\substack{+26 \\ - 28}$ & $5312\substack{+21 \\ -21}$& $5479\sus{16}{ 15}$ & $5489\sus{22}{ 21}$ \\
 \\[-1em]
 Spot temp., $\tau$ (K) & - & $2100\substack{+200 \\ -200}$ & - & -\\
 \\[-1em]
 Spot radius, $\rho$ (\textdegree) & - & $7.8\substack{+0.6 \\ -0.5}$  & - & -\\
 \\[-1em]
 Spot polar angle, $\theta$ (\textdegree) & - & $95\substack{+2 \\ -2}$ & - & -\\
 \\[-1em]
 Spot azimuth angle, $\phi$ (\textdegree) & - & $-27\substack{+1 \\ -1}$ & - & -\\
 \\[-1em]
 Diffusion coeff., $\kappa$ (W~K$^{-1}$~m$^{-2}$) & - & - & $-95\sus{7}{3}$ & 0\\
 \\[-1em]
 Diffusion index, $\Gamma$ & - & - & 0 & - \\
 \\[-1em]
 Convection amp., $v$ (J m$^{-2}$ K$^{-1}$) & - & - & $3230\sus{43}{45}$  & $3260\sus{22}{21}$ \\
 \\[-1em] 
  \hline
  \\[-1em]
 Volume-averaged $f$, $f_{\rm VA}$ & 0.952$\substack{+0.001 \\ -0.001}$ & 0.959$\substack{+0.001 \\ -0.001}$ & 0.965$\substack{+0.002 \\ -0.002}$ & 0.966$\substack{+0.001 \\ -0.001}$ \\
 \\[-1em]
 Effective radial velocity, $K_{\rm eff}$ (km s$^{-1}$) & $269\substack{+5 \\ -5}$ & $277 \substack{+4 \\ -5 }$ & $263\sus{3}{3}$ & $263\sus{5}{5}$\\
 \\[-1em]
  Distance, $d$ (kpc) & $1.64\substack{+0.09 \\ - 0.09}$ & $1.82\substack{+0.09 \\ -0.09}$ & $1.84\sus{0.06}{0.06}$ & $1.79\sus{0.1}{0.1}$\\
 \\[-1em]
 Pulsar mass, $M_{\rm psr}$ (M$_{\odot}$) & $0.93\substack{+0.04 \\ -0.04}$ & $1.13\substack{+0.05 \\ -0.04}$  & $0.96\sus{0.03}{0.03}$ & $0.96\sus{0.05}{0.04}$\\
 \\[-1em]
 Blackbody day temp. (K) & 6040 & 6170 & 6200 & 6210 \\
 \\[-1em]
 Blackbody night temp. (K) & 5310 & 5410 & 5430 & 5430 \\
 \\[-1em]
 \hline
 \\[-1em]
 Reduced chi-squared, $\chi^2_{\nu}$ & 2.37 & 1.11 & 1.38 & 1.38\\
 \\[-1em]
 Model evidence, $\log Z$ & -5177.6 & -3345.1 & -3906.4 & -3906.0\\
 \\[-1em]
 \hline
\end{tabular}
\caption{Numerical results for the modelling of J1227; including from top to bottom the model parameters, selected derived parameters, and model statistics. From left to right; standard symmetric model, single hot spot model, heat redistribution with linear diffusion and constant advection profile (HR1), and heat redistribution with convection only (HR2). $\log Z$ is the natural logarithm of the model evidence and $\chi^2_\nu$ is the reduced chi squared value with $\nu = 5908$ datapoints. Note that the inclination of the symmetrical model is higher than the others as the constraint described in section \ref{sec:J1227symresults} is not applied.}
\label{table:J1227}
\end{table*}
\end{center}

\begin{figure*}
	\includegraphics[width=0.99\textwidth]{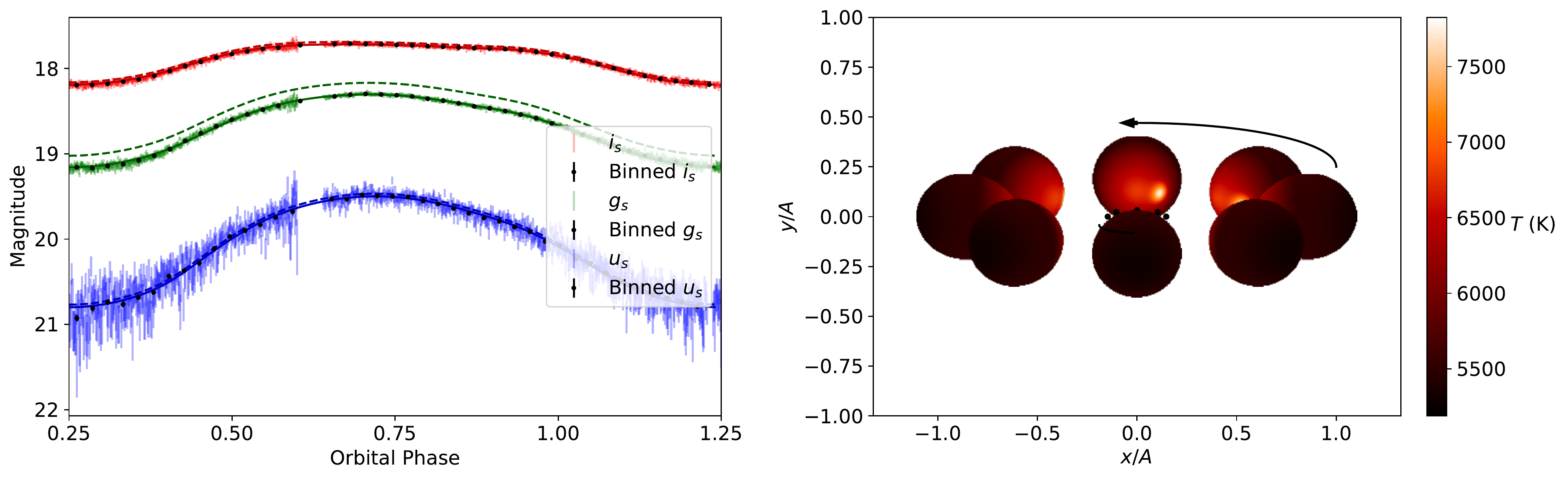}
    \caption{\textit{Left: }i$_{\rm s}$, g$_{\rm s}$, and u$_{\rm s}$\ light curves of J1227 overlaid with the best fit hot spot model. The dashed lines show the model light curve without band calibration corrections. \textit{Right: } Temperature distribution of companion - note how the inclination is such that the companion only just avoids eclipsing the pulsar at inferior conjunction.
    }
    \label{fig:J1227SpotFit}
\end{figure*}

\section{Results}
\label{sec:results}
\subsection{J1227}
\label{sec:J1227results}
\subsubsection{Standard model}
\label{sec:J1227symresults}
As expected, this model was unable to account for the asymmetries in the light curves, reflected by a reduced chi-squared value of $\chi^2_{\nu} = 2.37$ for our best-fitting model and evidence of -5177.6. We note that the evidence provides little information on its own, but is included as a means to compare each model. As a result, the model is a poor fit and the best-fit  parameters are likely erroneous; for example the sampler favours a nearly edge-on inclination of $i~=~88.3\substack{+0.7 \\ -1.0}$\textdegree\, converging at the upper limit of the prior. An inclination this close to 90\textdegree\ is highly unlikely as X-ray eclipses have not been observed \citep{deMartino2014}. In light of this, we include an additional constraint in subsequent fits; the pulsar (and hence the inner region of the accretion disc) must not be eclipsed by the companion star at any orbital phase. This results in an upper limit on the inclination, around $i_{\rm max}~\sim~77$\textdegree, though the exact value depends on the filling factor and mass ratio parameters.

The companion velocity is derived from the inclination and $K_2$ (and therefore $K_{\rm eff}$) in our model, such that the fit with velocity $K_{\rm eff}~=~269\substack{+5 \\ -5}$~km~s$^{-1}$, consistent with the \citet{deMartino2015} velocity, corresponds to an unrealistically low pulsar mass of $M_{\rm psr}\sim0.9{\rm M}_{\odot}$. Loosening our prior on $K_2$ results in a more acceptable pulsar mass of $M_{\rm psr}\sim1.2{\rm M}_{\odot}$  but a companion radial velocity of $K_{\rm eff}~=~308\substack{+12 \\ -20}$~km~s$^{-1}$, which is clearly not consistent with the literature. The companion temperature ($T~=~5452\substack{+19 \\ - 20}$ K) and irradiation temperature ($T_{\rm irr}~=~5230\substack{+26 \\ - 28}$ K) do agree with those determined in \citet{deMartino2015}, as these parameters are primarily influenced by the colour information rather than the shape of the light curves. However we do not reproduce their filled RL ($f~=~1.0$); instead we determine $f~=~0.825\substack{+0.002 \\ -0.002}$, corresponding to a volume-average filling factor of $f_{\rm VA} = 0.95$. Note also that their modelling uses a symmetric model. Considering these discrepancies in addition to a poor overall fit, we conclude that the standard Icarus heating model is not appropriate for this source. 

\subsubsection{Hot spot model}
The hot spot heating model was able to account for the majority of the asymmetry with $\chi^2_{\nu} = 1.11$ and evidence of -3345.1, indicating a significantly improved fit over the symmetric model. The best-fitting hot spot model is shown in figure \ref{fig:J1227SpotFit}. We find that the filling factor determined by our best-fit model, $f = 0.838 \pm 0.003$ ($f_{\rm VA} = 0.958 \pm 0.003$), also indicates that the companion is not Roche-lobe filling and is consistent with the symmetric model result. The system distance, $1.82\substack{+0.09 \\ -0.09}$~kpc and best-fit temperatures, $T_0 = 5556\substack{+14 \\ -15}$~K and $T_{\rm irr} = 5312\substack{+21 \\ -21}$~K, are also broadly similar to the symmetric model.

As with the direct heating model, the model prefers a $K_{\rm eff}$ comparable to that obtained from the spectroscopy, while the pulsar mass is unreasonably low, at $0.93 \pm 0.04 \ {\rm M}_{\odot}$. We therefore constrained the pulsar mass to the range $1.0 \text{M}_{\odot} < M_{\rm psr} < 3.0\text{M}_{\odot}$ for a repeat of this fit. However, the companion effective radial velocity with this constraint, $307 \substack{+10 \\ -10 }$~km~s$^{-1}$, is unacceptably large compared to the spectroscopic measurement of \citet{deMartino2014}. This occurs as $K_2$ increases to compensate for the high inclination, increasing $q$ as well.

\begin{figure*}
	\includegraphics[width=0.99\textwidth]{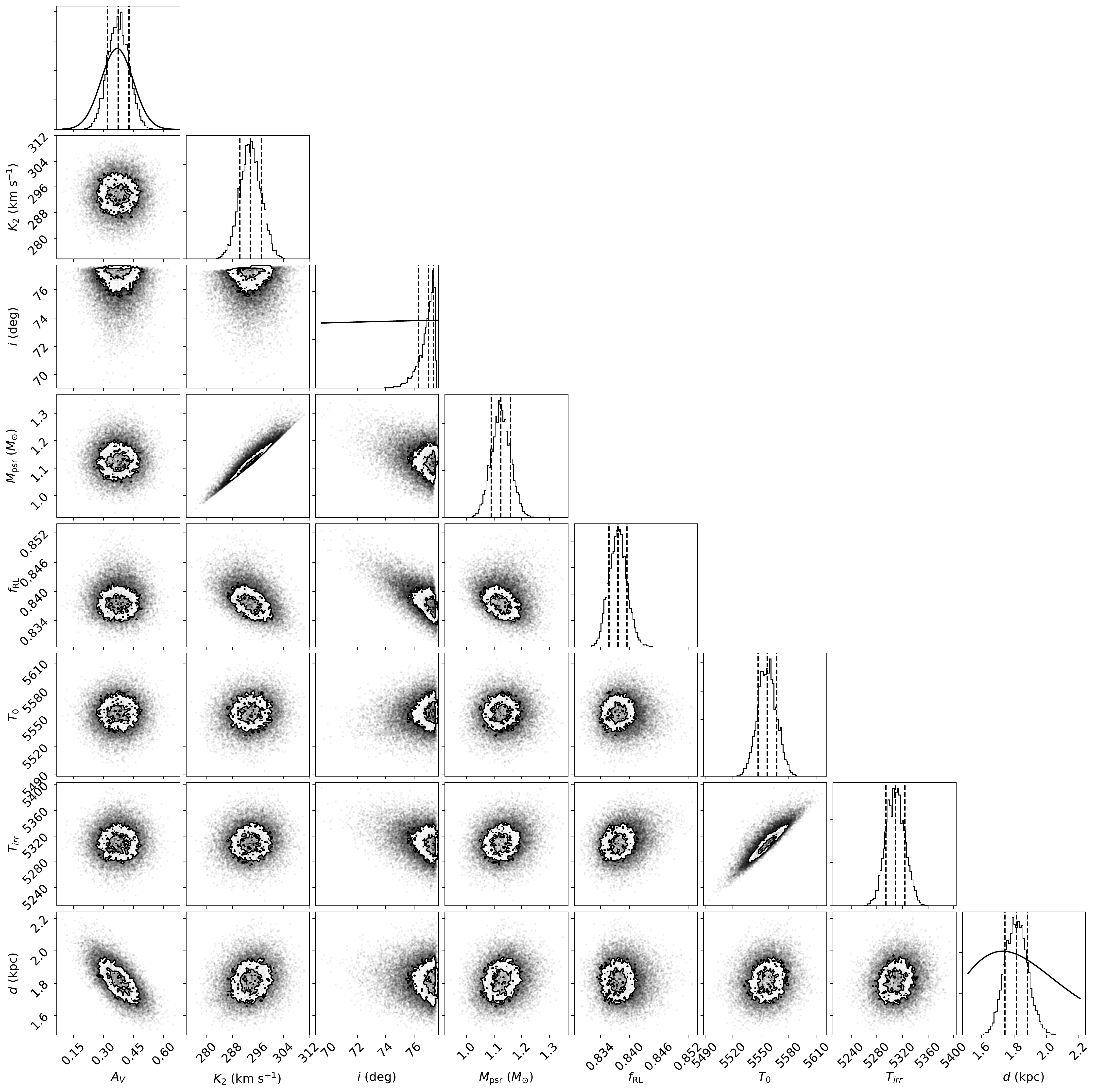}
    \caption{Corner plot of selected parameters of the J1227 hot spot model. Not shown are the hot spot parameters and the mass ratio. The plots along the diagonal are the posterior distributions of the parameters; the solid black lines on these are the prior distributions.
    The remaining plots show the position of walkers, illustrating covariance between parameters. Note the posterior distribution of the inclination, with the walkers converged against the upper limit set by the zero eclipse width.
    }
    \label{fig:J1227corner}
\end{figure*}

The inclination again indicates a nearly edge-on system with a best-fit value of $76.6 ^{+0.6}_{-1.2}$\textdegree. The shape of the posterior distribution is skewed, showing that the model has converged with an inclination very close to the limit imposed by the eclipse limit. This is illustrated in the corner plot in figure \ref{fig:J1227corner}. While the hot spot model provides an improved fit to the data, the best-fit values of the inclination, pulsar mass, and companion radial velocity are not suitable. 

\subsubsection{Cold Spots}
\label{sec:J1227coldspots}
The asymmetry of the light curve of J1227 is stronger in the cooler i$_{\rm s}$\  and g$_{\rm s}$\  bands than the u$_{\rm s}$\ band, suggesting that a cold spot may be better suited to model the asymmetry. We repeated the analysis in presented section \ref{sec:J1227fixed} using a cold spot model and found that the same trends were present. However, the distances and masses are larger than with the hot spot and for all inclinations the fitting is poorer. Notably, the effective $K_{\rm eff}$ velocities are consistently more than 1-sigma larger than the spectroscopic $K_2$, whereas these velocities were consistent when using the hot spot. Comparing the evidence for the $i=60$\textdegree\ run as an example, this was -3933.7 for the cold spot compared to -3600.8 for the hot spot, indicating a less favourable model. These factors indicate that the hot spot model is preferred over the cold spot for J1227.

\subsubsection{Heat redistribution}
We performed several fits using the heat redistribution model introduced in section \ref{sec:hr}, though none improved on the model parameters or evidence of the best-fitting hot spot model; that is, the edge-on inclination and small pulsar mass were still favoured. These results are presented in table \ref{table:J1227} alongside the results from the symmetric and single spot models. Note the similarity between the parameters of the two models, which results in indistinguishable model light curves. While the diffusion coefficient ($\kappa$) must be positive, we used a lower bound of -100 to avoid boundary effects around 0. As such, the negative value of diffusion coefficient in model HR1 (linear diffusion and convection) is unphysical and suggests a true value of 0. Additionally the uncertainty of this value is likely underestimated as the posterior distribution converged on the boundary. Indeed, the model parameters of HR1 and HR2 are otherwise consistent within uncertainties, suggesting that the model with convection only is a better description of the system. However, the model evidence and $\chi^2$ in both cases favours the hot spot model.

\subsubsection{Modelling assuming a filled Roche Lobe}
\label{sec:J1227fixedRL}
We also performed fits where the Roche Lobe is filled, with $f=1.0$ as in \citet{deMartino2014}. Under this assumption, the symmetric model provided an unsatisfactory fit similar to the $f$-free symmetric model and so was discarded. Modelling with a hot spot returned an acceptable fit with a evidence of -3871.8 and a reduced $\chi^2$ of 1.32; comparable but not better than the $f$-free case. However, the best-fit value of $K_2$ corresponds to a velocity of $K_{\rm eff} = 217 \pm 5$~km~s$^{-1}$, which is more than 8 standard deviations from the spectroscopic measurement.

\subsubsection{Modelling with fixed inclination}
\label{sec:J1227fixed}
With the inclination otherwise unconstrained, both the standard and hot spot models strongly favour an edge-on system with an inclination close to 90\textdegree. A system this edge-on is ruled out by the non-detection of X-ray eclipses, suggesting an inclination of $i\lesssim73$\textdegree\ \citep{deMartino2014}, and the lack of eclipses seen in spectra. Furthermore, when the radial velocity constraint from \citet{deMartino2014} is enforced the model returns a pulsar mass in the range $M_{\rm psr} \sim 0.8 - 1.0 {\rm M}_{\odot}$ which is clearly inappropriate. Conversely, relaxing this constraint we obtain a reasonable pulsar mass of $M_{\rm psr}\sim 1.4 {\rm M}_{\odot}$ but a value for $K_2$ which is too large, i.e. $\sim 330$~km~s$^{-1}$.

To attempt to overcome these discrepancies, we performed modelling of the system with the inclination fixed at each of $i = 40\text{\textdegree, } 50\text{\textdegree, } 60\text{\textdegree, } 70\text{\textdegree}$, using the hot spot model. Broadly, we observe that the pulsar mass, $T_0$, and filling factor are negatively correlated with inclination, while the system distance and irradiation temperature are positively correlated. The other parameters are not affected within uncertainties. We summarise these results in table \ref{table:inclination_results}, though we do not cover the full results of the fit with $i=40$\textdegree\ as the pulsar mass of $M_{\rm psr} \geq 3.0 {\rm M}_{\odot}$ and the distance of $d = 2.41\pm0.12$~kpc are both unreasonably large, suggesting that inclinations this low can be safely discarded.

It is worth noting that while the fit value of $K_2$ does not seem to be correlated with the inclination, the effective value appears to have a slight positive correlation with the inclination. However, the $K_{\rm eff}$ is consistent with the \citet{deMartino2014} radial velocity for inclinations 50\textdegree, 60\textdegree, and 70\textdegree, suggesting that the prior on the radial velocity is still tightly constraining. Likewise, the RL is consistently under-filled at all inclinations. We note that the evidence of each fit is also correlated with the inclination in a direction that suggests the model favours a more edge-on system, echoing what we observed when modelling with the inclination unconstrained: an edge-on inclination is favoured despite strong penalties from priors. We attempted modelling with the inclination tightly constrained rather than fixed, at $60\pm1$\textdegree, however the inclination did not converge to a solution after $\sim 5\times$ the usual computing time, suggesting this configuration is not appropriate. While this investigation shows the behaviour of the model at more face-on inclinations, the reason that the unconstrained inclination consistently converges to $i \sim 90$\textdegree\ is still unclear.

\begin{figure}
    \centering
    \includegraphics[width=0.49\textwidth]{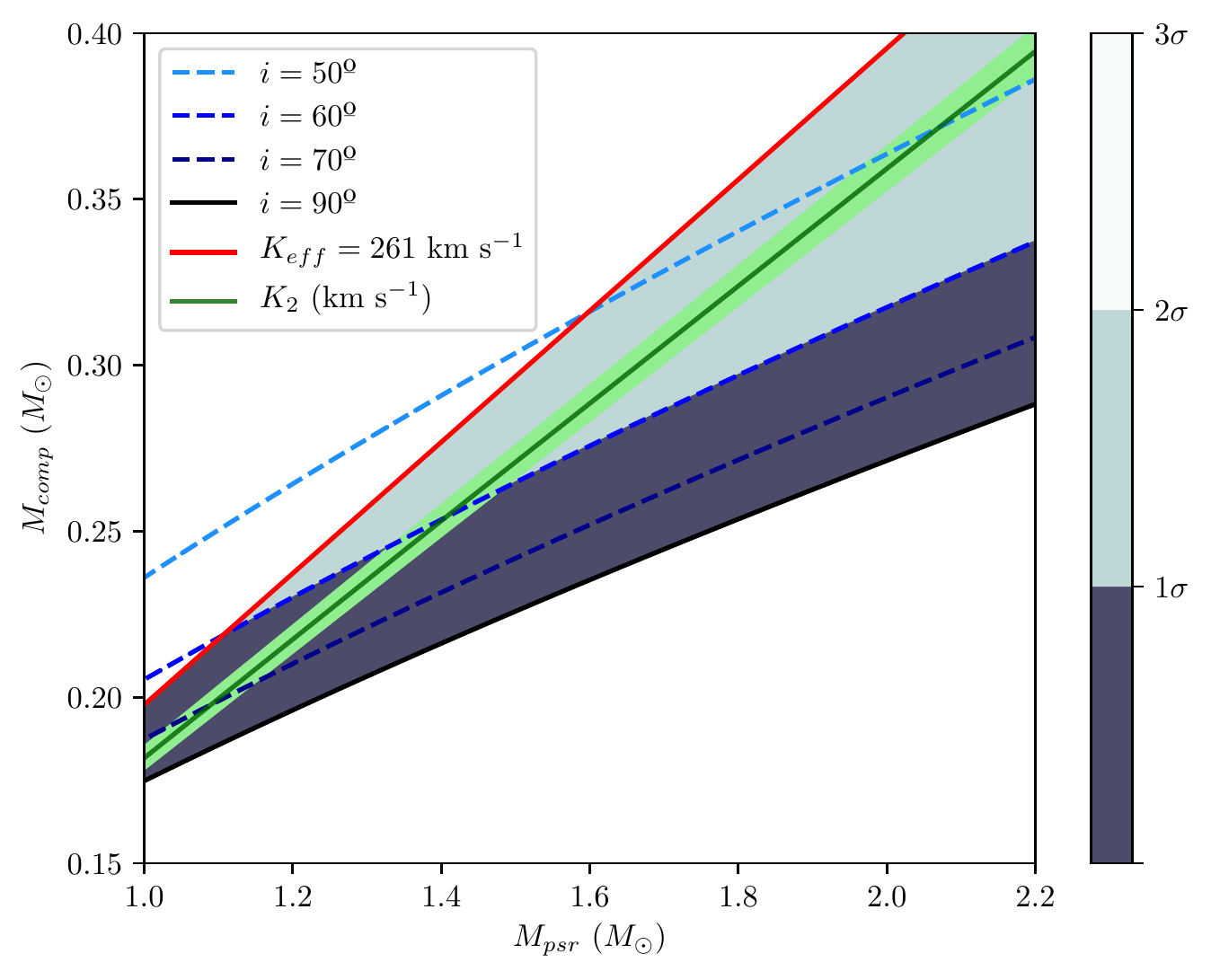}
    \caption{The mass ratio calculated from the centre-of-light $K_{\rm eff}$ value, in red, provides us with a lower limit of the pulsar mass for each inclination, while the best-fit mass ratio from each fit's $K_2$ is in green. The binary mass function at each inclination is shown by the series of blue dashed lines, with the absolute limit at $i=90$\textdegree\ shown by a solid black line. As such, the excluded region of the mass-mass plot is in white. The best-fit model distance at each inclination is interpolated over the mass-mass plane. The shaded area then represents the corresponding confidence interval of the distances with respect to the distance prior.}
    \label{fig:J1227-massmass}
\end{figure}

We constructed a mass-mass plot using the results in table \ref{table:inclination_results}, shown in figure \ref{fig:J1227-massmass}. This plot reveals constraints that we can apply to the masses of the companion and pulsar. We calculate a lower limit to the mass ratio, $q_{\rm min}$, from the centre-of-light $K_{\rm eff}$ radial velocity. Since in Icarus, the mass ratio is calculated from the larger, centre-of-mass $K_2$, this $q_{\rm min}$ acts as a lower bound to the pulsar mass at each inclination. These $K_2$ velocities at each inclination therefore correspond to the best fit mass ratio, and are consistent across the whole range to within the uncertainties. To further constrain the object masses, we interpolated the model distance estimates at each inclination in the allowed region of the plot. Then, by using the same prior distribution used during the fitting (a combination of GAIA parallax, galactic MSP density, and velocity distributions and the DM distance), we calculated the  confidence interval over the mass-mass plane. This shaded region indicates that the more edge-on inclinations produce more favourable distances. Indeed, these also correspond to better $\chi^2_{\nu}$ values and evidence. Using the evidence alone, the most favoured inclination is $i = 70$\textdegree, however this is a comparatively worse evidence than the best fitting free inclination model. However, we can conclude that for the range of inclinations $i\sim 50\text{\textdegree}-70\text{\textdegree}$, we obtain a pulsar mass in the range $M_p\sim 1.09{\rm M}_{\odot} - 2.0 {\rm M}_{\odot}$ and a companion mass in the range $M_c \sim 0.2 {\rm M}_{\odot} - 0.37{\rm M}_{\odot}$. The lower-bound pulsar mass of $M_{\rm psr} \approx 1.09 {\rm M}_{\odot}$ at $i=70$\textdegree\ would make J1227 the least massive known MSP and indeed very close to the lowest possible pulsar mass under current formation mechanisms. 

Since the mass estimates at more face-on inclinations more closely resemble those found in spiders, this is at odds with the better evidence and distance estimate at 70\textdegree. Considering these discrepancies we may surmise that the hot spot model is not a complete description of the system.

% Run numbers: M21 | M20 | M23
\begin{center}
\begin{table}
 \begin{tabular}{l  c  c  c } 
 \hline
 Model Parameters & $i=50$\textdegree & $i=60$\textdegree & $i=70$\textdegree \\ [0.5ex] 
 \hline\hline
 \\[-1em]
 $f$ & $0.901\substack{+0.02 \\ -0.02}$ & $0.866\substack{+0.002 \\ -0.002}$ & $0.841\substack{+0.002 \\ -0.002}$ \\
 \\[-1em]
 $M_{\rm psr}$ (M$_{\odot}$) & $2.1\substack{+0.1 \\ -0.1}$ & $1.41\substack{+0.06 \\ -0.06}$ & $1.09\substack{+0.05 \\ -0.05}$ \\
 \\[-1em]
 $d$ (kpc) & $2.1\substack{+0.1 \\ -0.1}$ & $1.8\substack{+0.1 \\ -0.1}$ & $1.67\substack{+0.09 \\ -0.09}$ \\
 \\[-1em]
 $T_0$ (K) & $5233\substack{+16 \\ -16}$ & $5288\substack{+15 \\ -17}$ & $5329\substack{+17 \\ -19}$ \\
 \\[-1em]
 $T_{\rm irr}$ (K) & $5356\substack{+26 \\ -26}$ & $5184\substack{+23 \\ -26}$ & $5080\substack{+26 \\ -28}$ \\
 \\[-1em]
 $K_2$ (km s$^{-1}$) & $282\substack{+5 \\ -5}$ & $280\substack{+5 \\ -5}$ & $280\substack{+5 \\ -5}$ \\
 \\[-1em]
 $K_{\rm eff}$ (km s$^{-1}$) & $258\substack{+5 \\ -5}$ & $262\substack{+5 \\ -5}$ & $264\substack{+5 \\ -5}$ \\
 \\[-1em]
 \hline
 \\[-1em]
 $\chi^2_{\nu}$ & 1.43 & 1.30 & 1.24 \\
 \\[-1em]
 $\log Z$ & -4099.9 & -3837.3 & -3690.3 \\ 
  \\[-1em]
 \hline
\end{tabular}
\caption{Model parameters and MultiNest evidence for fixed-inclination modelling of J1227 at $i=50\text{\textdegree, }60\text{\textdegree, }70\text{\textdegree}$. This model has $\nu=5908$ datapoints.}
 \label{table:inclination_results}
\end{table}
\end{center}

% %%%%%%%%%%%%%%%%%%%%%%%%%%%%%%%%%%%%%%%%%%%%%%%%%%%%%%%%%%%%%%%%%%%%%%%%%%%%%%%%%%%%%%%%%%%%%%%%%%%%%%%%%%%%%%%%%%%%%%%%%%%%%%%%%%%%%%%%%%%%%%%%%%%%%%
% %%      J1023
% %%%%%%%%%%%%%%%%%%%%%%%%%%%%%%%%%%%%%%%%%%%%%%%%%%%%%%%%%%%%%%%%%%%%%%%%%%%%%%%%%%%%%%%%%%%%%%%%%%%%%%%%%%%%%%%%%%%%%%%%%%%%%%%%%%%%%%%%%%%%%%%%%%%%%%
% Old Run numbers: M04 | M12 | N06 | N10
% new run numbers: F20 | F08 | F10 | F09
\begin{center}
\begin{table*}
 \begin{tabular}{l  c  c  c c} 
 \hline
 Parameters & Symmetrical & Single spot & HR 1 & HR2\\ [0.5ex] 
 \hline\hline
  \\[-1em]
 Inclination, $i$ (\textdegree) & $46.4\substack{+0.5 \\ -0.7}$ & $45.1\substack{+0.8 \\ -0.9}$ & $48.7 \substack{+ 0.20\\ -0.23 }$ & $45.3 \substack{+ 0.6\\ -0.5 }$\\ 
 \\[-1em]
 Mass ratio, $q$ & $7.8\substack{+0.05 \\ -0.05}$ & $7.8\substack{+0.1 \\ -0.1}$ & $7.87\substack{+ 0.04\\ -0.04 }$ & $7.89\substack{+ 0.05\\ -0.05 }$\\
 \\[-1em]
 Radial velocity, $K_{2}$ (km s$^{-1}$) & $295\substack{+3 \\ -3}$ & $295 \substack{+3 \\ -3 }$ & $297.0\sus{2.3}{ 2.2}$ & $298.0\sus{2.5}{ 2.5}$\\
 \\[-1em]
 Filling factor, $f$ & $0.86\substack{+0.01 \\ -0.01}$  & $0.94\substack{+0.02 \\ -0.02}$ &$0.808\sus{0.003}{ 0.003}$ & $0.864\sus{0.007}{ 0.007}$  \\
 \\[-1em]
 Base temp., $T_0$ (K) & $5580\substack{+14 \\ - 13}$ & $4885\substack{+31 \\ -30}$ & $4477 \sus{107}{43}$ & $5500\sus{10}{10}$\\
 \\[-1em]
 Irradiation temp., $T_{\rm irr}$ (K) & $4903\substack{+26 \\ - 23}$ & $4677\substack{+48 \\ -45}$ & $7351\sus{60}{150}$ & $4867\sus{15}{ 15}$\\
 \\[-1em]
 Spot temp., $\tau$ (K) & - & $1134\substack{+32 \\ -32}$ & - & -\\
 \\[-1em]
 Spot radius, $\rho$ (\textdegree) & - & $48\substack{+2 \\ -2}$  & - & -\\
 \\[-1em]
 Spot polar angle, $\theta$ (\textdegree) & - & $8.2\substack{+1.0 \\ -0.8}$ & - & -\\
 \\[-1em]
 Spot azimuth angle, $\phi$ (\textdegree) & - & $-57\substack{+5 \\ -5}$ & - & -\\
 \\[-1em]
 Diffusion coeff., $\kappa$ (W~K$^{-1}$~m$^{-2}$) & - & - & $93500\sus{4000}{8000}$ & 0\\
 \\[-1em]
 Diffusion index, $\Gamma$ & - & - & 0 & - \\
 \\[-1em]
 Convection amp., $v$ (J m$^{-2}$ K$^{-1}$) & - & - & $52000\sus{2000}{4000}$  & $8630\sus{60}{60}$ \\
 \\[-1em]
 \hline
 \\[-1em]
 Volume-averaged $f$, $f_{\rm VA}$ & 0.969$\substack{+0.009 \\ -0.008}$ & 0.994$\substack{+0.004 \\ -0.003}$ & 0.942$\substack{+0.002 \\ -0.002}$ & 0.971$\substack{+0.004 \\ -0.004}$ \\
 \\[-1em]
 Effective radial velocity, $K_{\rm eff}$ (km s$^{-1}$) & $287\substack{+3 \\ -3}$ & $283 \substack{+3 \\ -3 }$ & $291.0\sus{2.3}{ 2.2}$ & $290.0\sus{2.5}{ 2.5}$\\
 \\[-1em]
  Distance, $d$ (kpc) & $1.26\substack{+0.02 \\ - 0.01}$ & $1.28\substack{+0.02 \\ -0.02}$ & $1.19\sus{0.01}{0.01}$ & $1.30\sus{0.02}{0.02}$\\
 \\[-1em]
 Pulsar mass, $M_{\rm psr}$ (M$_{\odot}$) & $1.76\substack{+0.06 \\ -0.05}$ & $1.89\substack{+0.10 \\ -0.09}$ & $1.69\sus{0.03}{0.03}$ & $1.62\sus{0.06}{0.06}$\\
 \\[-1em]
 Blackbody day temp. (K) & 5820 & 5700 & 5750 & 5750 \\
 \\[-1em]
 Blackbody night temp. (K) & 5470 & 5380 & 5420 & 5410 \\
 \\[-1em]
 \hline
 \\[-1em]
 $\chi^2_{\nu}$ & 6.20 & 1.20 & 1.26 & 1.45\\
 \\[-1em]
 $\log Z$ & -5295.0 & -2293.9 & -2462.6 & -2633.5\\
 \\[-1em]
 \hline
\end{tabular}
\caption{Numerical results for the modelling of J1023; including from top to bottom the model parameters, selected derived parameters, and model statistics. This model has $\nu=3821$ datapoints. From left to right; standard symmetric model, single hot spot model, heat redistribution with linear diffusion and constant advection profile (HR1), and heat redistribution with convection only (HR2). Note the similarity between the blackbody temperatures of each model, despite the large range of $T_0$ and $T_{\rm irr}$ temperatures. }
\label{table:J1023table}
\end{table*}
\end{center}

\subsection{J1023}
\label{sec:J1023results}
\subsubsection{Standard Model}
\label{sec:J1023symresults}
As with J1227, the symmetric heating model did not provide a good fit to the data, with a best-fit reduced $\chi^2$ value of $\chi^2_{\nu} = 6.20$ ($\nu = 3821$) and evidence of -5295.0. This is a comparatively worse fit than the same modelling of J1227, in part because smaller error bars from a brighter source and better observing conditions make the asymmetry more significant compared to the noise, with the u$_{\rm s}$-band fit especially poorly.
The best-fit parameters are summarised in table \ref{table:J1023table}.

We determine an inclination angle of $i = 46.4 \sus{0.5}{0.7}$\textdegree, which is consistent with the range of possible inclinations of \citet{Archibald2009}, 34\textdegree\ $< i < 53$\textdegree, and consistent with the inference in \citet{Thorstensen2005} that $i < 55$\textdegree. Note that these literature values were calculated under the assumption that the companion is Roche lobe-filling. We again obtain a RL filling factor of significantly less than 1.0 though there is a strong negative correlation with the inclination, such that at an inclination consistent with the \citet{Archibald2009} calculation, the filling factor approaches $f=1.0$.

The system temperatures indicate that the companion is not as strongly irradiated as in J1227. A best-fit distance of $d = 1.26 \sus{0.02}{0.01}$ kpc broadly agrees with the radio parallax measurement of $1.368 \sus{0.042}{0.039}$~kpc from \citet{Deller2012}. Note that across all models the distance is consistently underestimated; this is discussed in section \ref{sec:J1023disc}. This fit produces a radial velocity of $K_{\rm eff} = 287 \pm 3$~km~s$^{-1}$, corresponding to a mass ratio of $q = 7.8 \pm 0.05$ which is comparable but not consistent with the result from radio timing in \cite{Archibald2009}. This is as expected, as the radial velocity parameter appears to be consistent with the upper bound of measurements from the spectral lines in \citet{Shahbaz2019}, and the mass ratio is calculated directly from this velocity. Unlike with our modelling of J1227 the posterior distributions of these model parameters, notably the mass, generally converged within the range of expected literature values. However, the large $\chi^2$ value leads us to again conclude that the symmetric model is insufficient.

\begin{figure*}
	\includegraphics[width=0.99\textwidth]{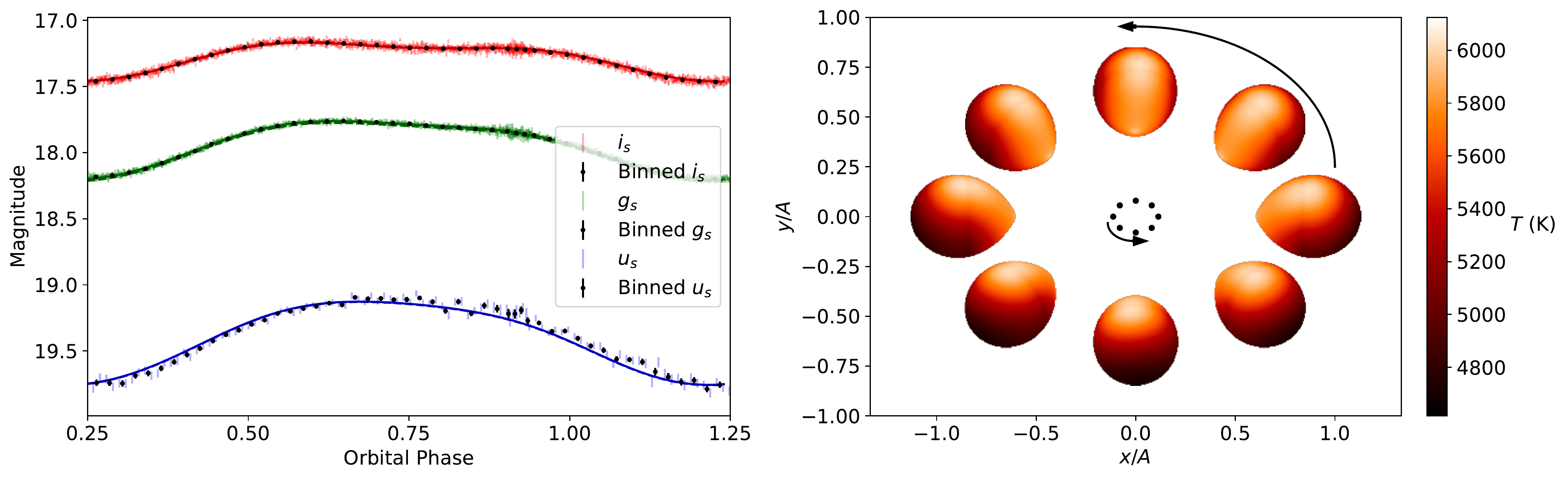}
    \caption{\textit{Left: }i$_{\rm s}$, g$_{\rm s}$, and u$_{\rm s}$\ light curves of J1023 overlaid with the best fit hot spot model, with 40 phase bins in black. \textit{Right: } Temperature distribution of companion, showing the asymmetry which manifests as a large, polar spot. The dark band around the star is an artefact due to the plotting only.
    }
    \label{fig:J1023-spot-model}
\end{figure*}

\subsubsection{Hot spot model}
\label{sec:J1023spotresults}
With $\chi^2_{\nu} = 1.20$ and evidence of -2293.9, the hot spot model provides a much better fit to the data than the symmetric model, capturing the asymmetries to a good degree. The parameters of the best-fitting model are shown in table \ref{table:J1023table}. The filling factor remains below 1.0 and the inclination is again consistent with that inferred from radio timing from \citet{Archibald2009}. The distance is also broadly consistent with the interferometry distance in \citet{Deller2012}; $1.28\pm 0.02$~kpc compared to $1.368 \sus{42}{39}$~kpc. Considering the best-fit spot parameters, the spot radius of $48 \pm 2$\textdegree\ and position near the companion pole is strikingly similar to the spot properties seen in \citet{Kandel2020}. The temperature distribution of the companion is shown alongside the best-fitting model in figure \ref{fig:J1023-spot-model}. We expect tMSPs to host massive neutron stars, and the constrained mass of $M_{\rm psr} =1.89\substack{+0.10 \\ -0.09}$ is no exception. Note however that there is a moderate inconsistency between the determined pulsar masses.

\subsubsection{Heat Redistribution}
Choosing a diffusion index of 0 (linear diffusion) and a constant advection polar convection profile, we ran an initial model of the light curve. We found that the asymmetry was well-fit, but with a marginally worse evidence and reduced $\chi^2$ than the best fitting hot spot model. Since the heat redistribution model also allows for a range of combinations of diffusion and convection, the results of \textit{(1)} a model with convection only, and \textit{(2)} convection with linear diffusion (i.e. with a zero diffusion index) are shown in table~\ref{table:J1023table}. However, model \textit{(3)}, with convection and temperature-dependent diffusion converged to a solution with a very large diffusion index which caused significant aliasing in the temperature distribution of the companion; as such these results have not been included. Comparing the other two models, the evidence favours a model with convection and linear diffusion, despite the fact that the best-fit distance for this model is significantly below the literature value.

\section{Discussion}
\label{sec:discussion}
\subsection{Filling factor}
\label{sec:discussionfilling}
One consistency across all models is an underfilled Roche lobe, with our modelling returning values in the range $f \sim 0.825 - 0.90$ for J1227 and $f\sim 0.81-0.94$ for J1023. This is at odds with our expectations for these sources; as they are both tMSPs we would expect the Roche lobes to be full or nearly filled in the rotation-powered (RP) states given they have transitioned from Roche lobe overflow in their accretion-powered (AP) states. However, when considering the volume-averaged fill factor, the values of $f_{\rm VA} = 0.958$ and 0.994 for the best-fitting models of J1227 and J1023 respectively tell a story more consistent with our initial expectations; that the Roche lobe is indeed mostly full. 

\begin{figure}
    \centering
    \includegraphics[width=0.38\textwidth]{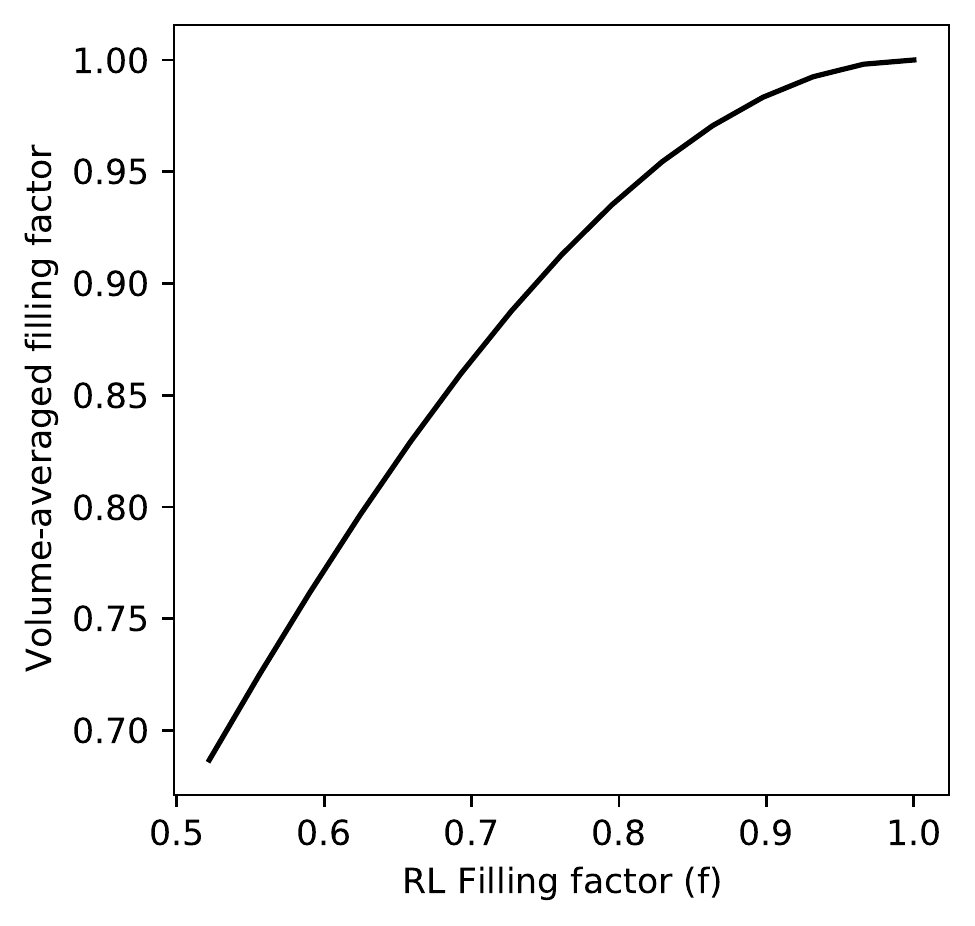}
    \caption{Relationship between filling factor, $f$, and the volume-averaged filling factor. At lower values of $f$ ($f < 0.6$) the relationship is linear. Considering filling factors for $f>0.8$, it can be seen how an apparently underfull Roche lobe can have a volume-averaged filling factor much closer to unity due to the tidally distorted shape of the companion.}
    \label{fig:vaf}
\end{figure}

Figure \ref{fig:vaf} illustrates the relationship between these two parameterisations. When considering the proximity of the companion star to a state of RL overflow, $f$ is a more useful description as it gives the distance of the star surface to the L1 point. However, the volume-averaged filling factor gives a clearer picture of the size of the star relative to its RL, in these cases illustrating that a substantial increase in the volume-averaged radius of the star is not necessarily required for RL overflow to begin. For J1023 in particular, the volume-averaged fill factor approaches unity. With a nearly-full Roche lobe in the RP state, we may consider mechanisms for transitions.

We may assume that if the mass transfer is conservative, then the orbital separation should increase during the AP state in order to conserve angular momentum. This would correspond with an expansion of the companion's Roche lobe. While orbital period variations in the RP state have been observed \citep{Archibald2013}, these are over the timescale of $\sim100$ days, shorter than the timescale between transitions of $\sim 1-10$ years. Furthermore, these variations are not of sufficient magnitude to cause the required changes in Roche lobe radius. The observed variation in $T_{\rm asc}$ are of order $\sim 1$ second. With the relation $\Delta T_{\rm asc} \sim \dot{P}_{\rm orb} t_{\rm obs}$, where $t_{\rm obs}\sim~100$~d is the observation time, this corresponds to a change in $a$ of $\Delta a / a \sim \num{e-5} $, a factor of $\sim~10^5$ too small. We assume that a fraction change in $a$ is equivalent to the same fractional change in the Roche lobe radius.

Lastly, there appear to be no correlated changes in the orbital period over the several years of data, suggesting the changes are not gradual and continuous as would be the case with steady mass loss from ablation. Instead, a change in the structure of the star without significant mass loss could be explained by the size of the convective envelope decreasing or disappearing completely as the system transitions from the AP to RP state. We assume that the companion stars in redback systems have large convective envelopes, so it is possible that these are `puffed up' while in the RP state.

The irradiation of the companion by the pulsar wind is known to expand the companion photosphere, however it is uncertain if this could occur sufficiently within the transition timescale. Similarly, if an accretion disc shields the secondary from the pulsar irradiation, it could allow the companion photosphere to gradually contract during the AP state. Consider the Kelvin-Helmholtz mechanism,
\begin{equation}
    \label{eq:KH}
    \tau_{\rm KH} \sim \frac{GM^2}{2RL} \sim \frac{2G\rho M}{3 \sigma_{\rm SB} T^4},
\end{equation}
where $G$ is the gravitational constant, $M$ is the mass of the star, $R$ is its radius, $L$ is its luminosity, $\rho$ is the mean density of the star, $T$ is its temperature, and $\sigma_{\rm SB}$ is the Stefan-Boltzmann constant \citep{KW2012}. Assuming a mean stellar density of 1~g~cm$^{-3}$, a companion mass of $0.4~{\rm M}_{\odot}$, and a temperature of 5500~K, we obtain a timescale of $\tau_{\rm KH} \sim \num{2e7}$~yr. This is significantly longer than the $\sim$~yr timescales between tMSP transitions. If we consider the contraction of only the outer convective layer, some fraction $k_R$ of the stellar radius, it is possible a thin layer of the companion could contract and expand within transition timescales. However, this is highly dependent on the depth of this necessarily thin layer.

In light of this, we consider the discussion of the envelopes of asymptotic giant branch (AGB) stars in \citet{Soker2015}. Following equation 2 in \citet{Soker2015}, we separate the companion star into a core of mass $M_{\rm core} = f_c M_c$ and envelope $M_{\rm env} = (1-f_c) M_c$, where $f_c$ is a fraction between 0 and 1. Assuming the radius of the star is equal to the Roche lobe radius, calculated using the Eggleton approximation \citep{Eggleton} with $q=7.8$, we calculate the Kelvin-Helmholtz  timescale for this envelope to be $\tau_{\rm KH} \sim f_c (1 - f_c) \times 100$~kyr. To obtain a timescale of order $\sim 10$~yr, as expected from the tMSP transition timescale, we arrive at a mass fraction of $f_c\sim \num{e-4}$. This is a plausible result, suggesting that a fraction of the envelope can be expected to contract within tMSP transition timescales.

Considering the corner plot in figure \ref{fig:J1227corner}, showing the J1227 hot spot model parameters, strong covariance between the filling factor and inclination can be seen. This covariance indicates a negative correlation, such that for a lower (more face-on) inclination, the filling factor would increase towards unity. Given that we suspect the inclination of J1227 to be over-estimated in our model, this may suggest that the star may be even closer to filling its Roche lobe than inferred. A similar relationship between the distance and filling factor is seen for J1023, where a distance closer to the interferometry measurement results in a filling factor closer to unity.

\subsection{Asymmetries}
\label{sec:asymm_disc}
\begin{figure}
    \centering
    \includegraphics[width=0.47\textwidth]{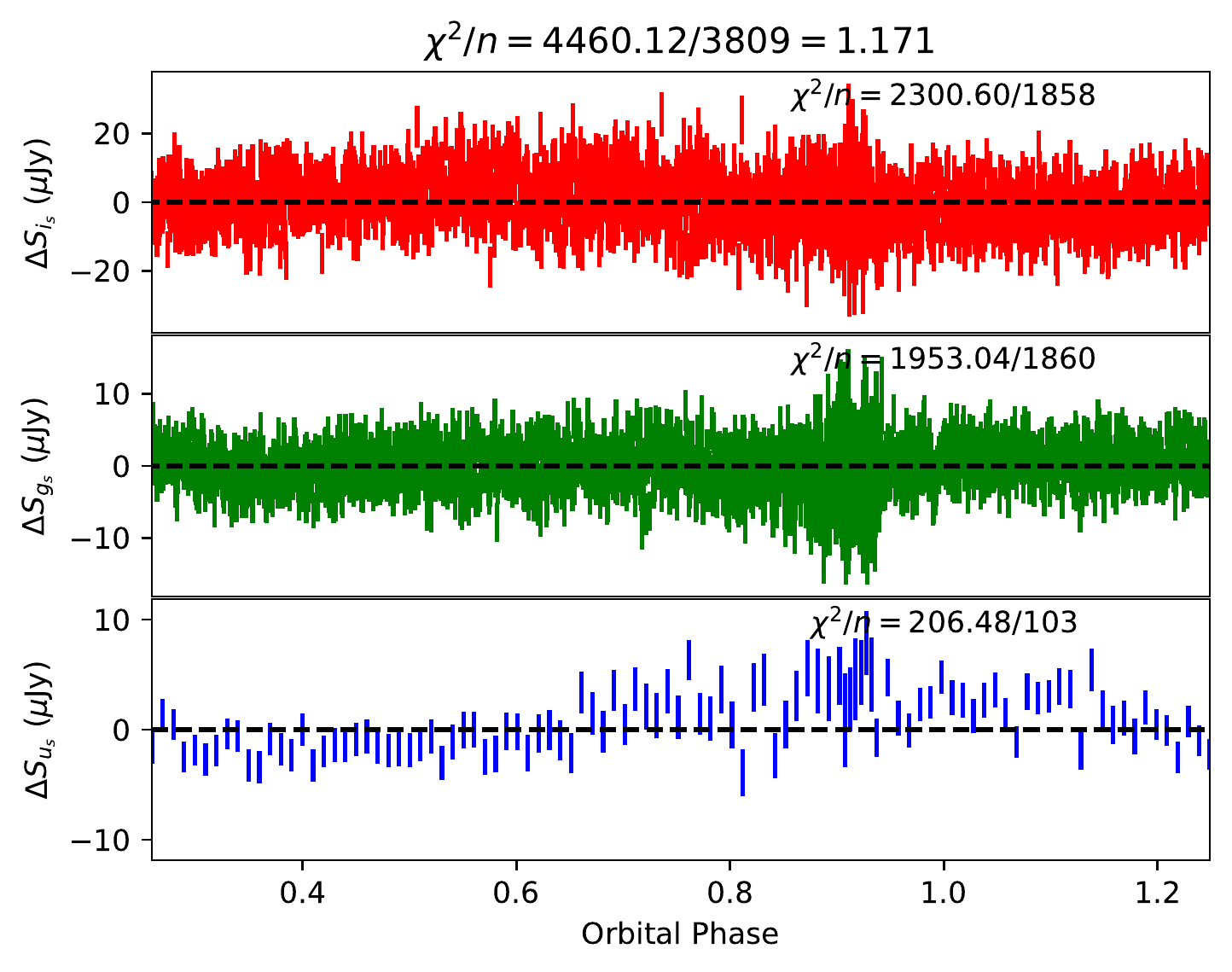}
    \caption{Residuals per band for the best-fitting hot spot model of J1023; note that the $y$-axis is in units of flux. The u' residuals show a clear systematic pattern that mirrors the sinusoidal shape of the asymmetry in figure \ref{fig:symmetriesJ1023}. Note also that the reduced $\chi^2$ value for the u' is significantly larger than that for the other two channels; 1.88 compared to 1.05 for g' and 1.31 for i'. }
    \label{fig:J1023M18res}
\end{figure}

For both sources the hot spot model clearly provides a better fit than the symmetric model \textemdash\ both in comparing the $\chi^2_{\nu}$ values, the Bayesian evidence, and the model parameters \textemdash\ but it is clear that it is still not a complete description of systems with asymmetric light curves. This is evidenced by, for example, the model favouring an edge-on inclination for J1227 in all situations, despite strong penalties from priors. Further to this, for both sources the residuals in the u$_{\rm s}$-band show systematic variations, implying that some aspects of the light curves are not captured by the model. The per-band residuals for the best-fitting hot spot model of J1023 are shown in figure \ref{fig:J1023M18res}. Examining these in more detail we find the model produces too much asymmetry in the u$_{\rm s}$ band compared to the data, while for the i$_{\rm s}$ and g$_{\rm s}$ bands the residuals show that the model is capturing the asymmetry well. Why this happens is not immediately clear. While it may be interpreted as further evidence that the hot spot model is insufficient, similar systematics are also present in the best-fitting heat redistribution model. 

\begin{figure}
    \centering
    \includegraphics[width=0.49\textwidth]{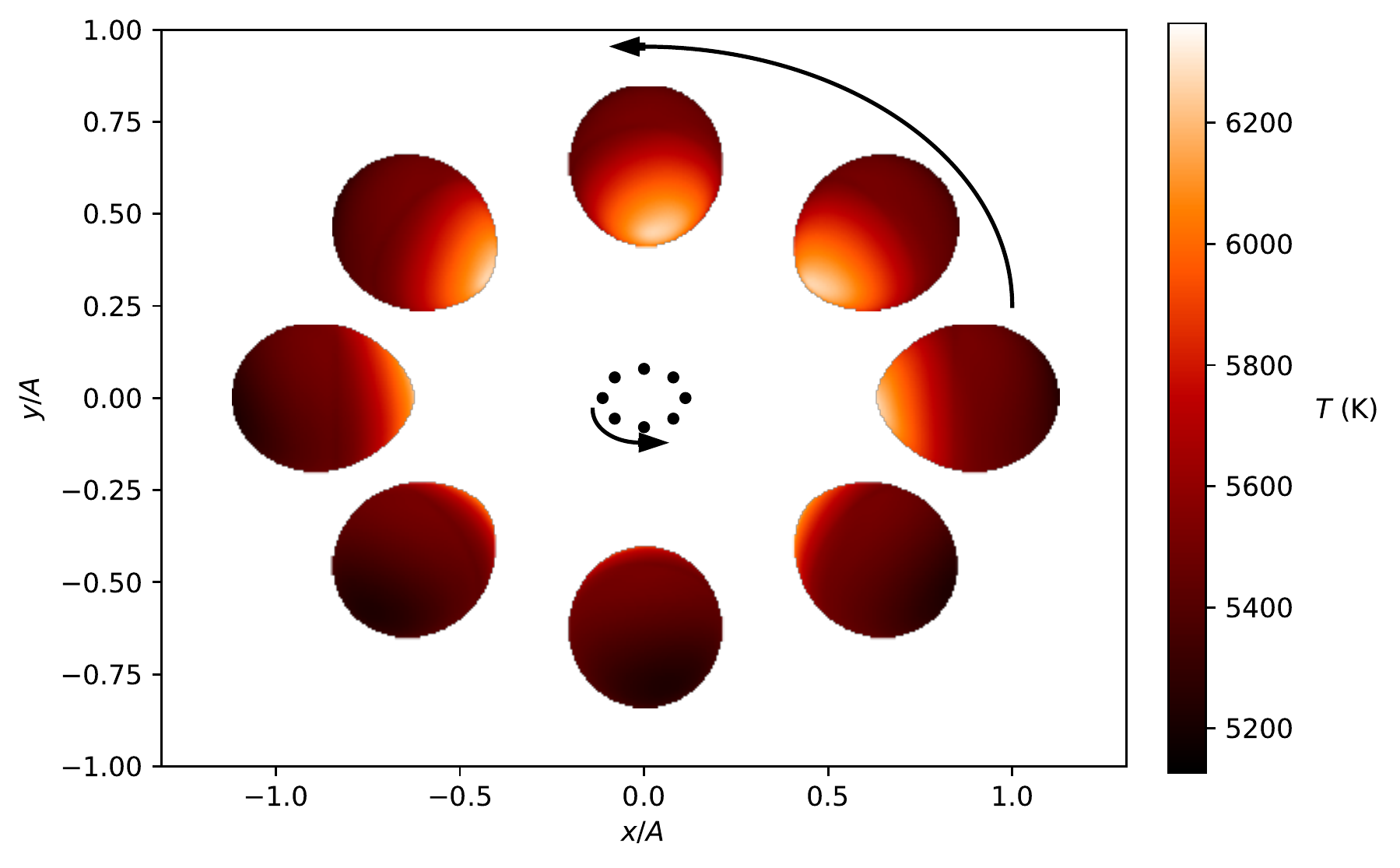}
    \caption{Companion star temperature distribution for the best-fit heat redistribution (HR1) model of J1023. The asymmetry of this distribution can be seen near the L1 point, and the lack of the polar spot seen in the hot spot model is no longer present. The model light curve and residuals are indistinguishable from the hot spot case.}
    \label{fig:J1023_HR_model}
\end{figure}

As is described in section \ref{sec:hr}, our heat redistribution model incorporates the effects of diffusion and convection in the companion photosphere. Our results using this model are of comparable quality to the hot spot model, the Bayesian evidence indicating an improved fit over the direct heating model. However, the temperature distributions obtained differ significantly from those in the hot spot model, for example for J1023 for which the polar spot disappears in the heat redistribution model. The temperature distribution of the best-fit heat redistribution model is shown in figure \ref{fig:J1023_HR_model}. A possible model not tested in this work  may be a combination of the two models; a hot spot with heat redistribution.

\subsection{J1227}
\label{sec:J1227disc}
Despite the modelling failing to provide a single best fitting solution, we can discuss several key findings. Our inclination range of $i~\sim~50$\textdegree$-70$\textdegree\ obtained by fitting at fixed inclinations agrees with the range proposed in \citet{deMartino2014, deMartino2015}, though these do not translate into a particularly strong mass constraint; a pulsar mass in the range $M_p\sim 1.09{\rm M}_{\odot} - 2.0 {\rm M}_{\odot}$ and a companion mass in the range $M_c \sim 0.2 {\rm M}_{\odot} - 0.37{\rm M}_{\odot}$. However, the tendency of the model towards edge-on inclinations is concerning.

The lack of observed X-ray eclipses and the fact that no eclipse is observed in the spectra argue against this edge-on inclination. The preference for a high inclination in the model suggests that the model is attempting to increase the fraction of ellipsoidal modulation relative to the irradiation. The amplitude of the ellipsoidal modulation is also proportional to the filling factor cubed,
\begin{equation}
    A_{\rm el} \sim f^3 q \sin^2i
\end{equation}
while the irradiation amplitude is proportional to the square of the filling factor \citep{Breton2012}. The high inclination may be compensating for a smaller filling factor, which may be caused by the model attempting to fit the asymmetries with a larger ellipsoidal term than is truly present. The source of asymmetry in the system is still uncertain; while the two proposed model extensions do offer a significantly improved fit the discrepancies between the model parameters and observables, notably the inclination, suggest that there is still significant physics in the system that is not understood.

Before performing the modelling with fixed inclination described in section \ref{sec:J1227fixed} we attempted to model the system with the pulsar mass fixed over a range of values, as with J1023. 
However, the $K_2$ velocities obtained with these models were unacceptably large; more than three standard deviations above the spectroscopic distances. 
While we might expect some systematic error in this spectroscopic value, it cannot be large enough to explain this discrepancy.
Since in these fixed-mass models the mass ratio and inclination are derived from $K_2$, the high inclinations that the model consistently prefers necessitate high $K_2$ velocities. 
With the inclination fixed, $q$ and $M_{\rm psr}$ are derived so this is no longer an issue and the model is well-behaved. 

\subsection{J1023}
\label{sec:J1023disc}
While several system parameters - namely the inclination, mass ratio, radial velocity, and filling factor - are broadly similar across the different models, some considerable differences remain. The temperature distributions and pulsar masses were starkly different in each case. Comparing the temperature distributions of the companion surfaces of each model we see that the large, polar spot seen on the hot spot model is not reproduced by the heat redistribution model, despite both models having very similar fit residuals. This could suggest that the large spot may be a sign of over-fitting, which would indicate that the magnetic ducting theory proposed in \citet{Kandel2020} is not appropriate here. However, many cataclysmic variables (CVs) and rotating stars also display polar spots \citep{Watson2007}, which may suggest that the similar alignment of magnetic fields is not coincidental. It should be noted that the spots in \citet{Watson2007} are cold spots, as opposed to the hot spots which are favoured in this work. An additional explanation may be drawn from the fact that the viewing angle of this polar spot does not change much over the orbit, suggesting that this spot configuration is in fact fitting for an additional, constant source of flux. 

\begin{figure*}
	\includegraphics[width=0.99\textwidth]{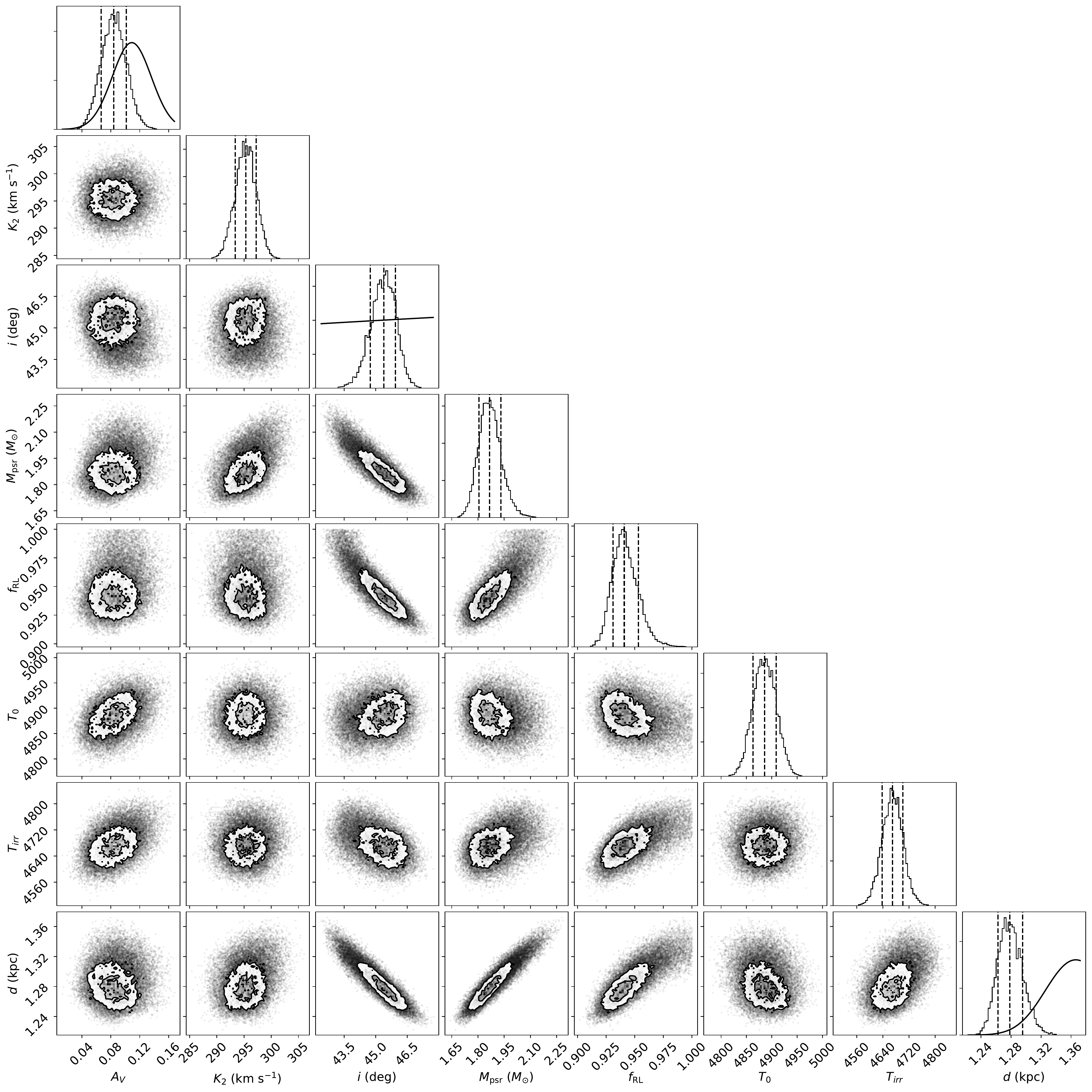}
    \caption{Corner plot of selected parameters of the J1023 hot spot model. Not shown are the hot spot parameters and the mass ratio. Note the discrepancy between the distance posterior and prior distribution, and the strong covariance of the distance with the inclination, pulsar mass, and filling factor.
    }
    \label{fig:J1023_corner}
\end{figure*}

As such, we introduce a power law with an index of unity in order to mimic a third light in the system similar to what is observed in the AP state from broadband spectroscopy (Hernandez Santisebastien, et al., private communication). This light is incorporated as a fraction, $t$, of the expected flux from the companion alone such that $t  = F_{\rm 3rd\ light} / F_{\rm companion}$. For each band, these correspond to $t_i = 0.150$, $t_g = 0.196$, and $t_u=0.625$, with the contribution from the third light most significant in the u$_{\rm s}$-band. We then adjust the fluxes prior to fitting by multiplying each band by the ratio $1 / (1 + t)$, such that the modelled contribution of the third light is removed and the remaining flux represents only the light from the companion. We use a heat redistribution model with linear diffusion and constant advection profile. We chose this model over the better-fitting hot spot model as for that model the free temperature of the spot makes it difficult to clearly separate the flux contributions of each source. That is, the spot temperature can easily decrease to compensate for the reduced flux.

While the adjusted i$_{\rm s}$\ and g$_{\rm s}$\ bands were fit well by the model, the adjusted flux of the u$_{\rm s}$\ band could not be matched. This suggests that either our power law model does not accurately describe the third light flux, or that a third light is not able to account for the distance discrepancy. Notably, the shape of the residuals in u' were unchanged compared to a fit using unaltered fluxes. We note that the spectra presented in \citet{Shahbaz2019} show no evidence for a stellar third light. A continuum emission source such as the synchrotron emission produced in the intrabinary shock \citep{Romani2016} may be an alternative source of the flux, however this emission follows a negative power law. Shahbaz et al. (in prep. and private communication) further show no evidence for a third light, with the secondary star the sole source of flux from 6000~\AA. However, a metal-rich secondary is observed, with an iron excess of Fe/H = 0.48 (\citealt{Shahbaz2019} and private communication). This results in a u'-band flux of 82~per~cent the solar equivalent in our model, which may account for the closer distance that we obtain due to excess flux; a metal rich star is less blue compared to solar. A system distance of between 1.25 and 1.30 kpc, as we obtain, compared to the interferometry distance of 1.368 kpc corresponds with a decrease in magnitude of between 0.11 and 0.2 mag in the u$_{\rm s}$ band. This is equivalent to a decrease in flux of $83-90$~per~cent, which is comparable to the expected decrease due to a higher metallicity. The atmosphere grids used in this modelling do not account for this high iron excess, and as such this avenue may help to explain the distance we obtain.

While the best fitting values are different for the heat redistribution and hot spot models, both show a strong covariance between the filling factor, $f$, and the distance. This can be seen in the corner plot shown in figure \ref{fig:J1023_corner}. This suggests that a better fit to the distance will bring the filling factor closer to unity, as we would initially expect.  

Considering the competing prior values for the $A_V$ extinction, our modelling shows that both values are almost equally favoured as no other parameters are affected by the change of prior within uncertainties. As such, we can surmise that the small difference in model evidence (-2297.1 for the value from $N_H$ and -2293.9 for the $A_V$ from dust maps) are due only to the differences in penalty from the $A_V$ priors. The $A_V$ value from the dust maps, which provides the marginally improved model evidence, was used for the modelling in this work, though in practice this investigation shows that the model parameter posteriors are nearly ideally distributed and any changes to the $A_V$ have little effect on the results.

\section{Conclusions}
\label{sec:conclusion}
We present new, high time resolution optical photometry of the tMSPs PSR J1023+0038 and PSR J1227$-$4853, and discuss our numerical modelling of their light curves. Using a new extension to the Icarus code including the thermal contributions of a hot or cold spot on the companion surface, we modelled the asymmetric light curves of these sources and obtained significantly improved fits over the symmetric case. Using this model we constrained several key parameters of the systems, including the companion Roche lobe filling factor, temperature profile, and system distance. We also performed modelling using a further extension which considers the diffusion and convection on the companion star surface. This model also provided an improved fit compared to the symmetric model, though for both systems the evidence favours a model with a hot spot.

We found that the filling factor of both sources was less than 1.0, indicating that the Roche lobe is under-filled. This is at odds with other results showing a full Roche lobe in the system's RP state \citep{deMartino2014}. We expect the companion in the AP state to fill its Roche lobe, so these results indicate that the filling factor plays an important role in the tMSP transition. However, when considering the volume-averaged filling factor, we find that the companion stars are only slightly underfilling their Roche lobes. Our results suggest that the companion stars may undergo an expansion and contraction between the AP and RP states of tMSP cycle, as there is no sign of a change in the size of the Roche lobe; changes in the orbital period are not large enough to account for the under full Roche Lobes. Taking our filling factors at face value, they indicate that between state transitions the stellar radius changes by an order or 5~per~cent, something which may be possible if the companion has a core-envelope structure with only a small fraction of its mass in the envelope.

Some of the limitations and errors in our results indicate that our model does not completely describe the asymmetry of the light curves, suggesting that improvements to the model are needed, or indeed a new, physically motivated approach. For example, significant systematics remain in the residuals of the u$_{\rm s}$ band of J1023. Using a further extension to the Icarus code which models diffusion and convection in the companion photosphere we performed a second round of modelling. This again provided an improved fit over the symmetric model, however the inconsistent results, such as the inclination of J1227, persisted, and indeed the fit overall was poorer than the hot spot case. However, modelling with some combination of the two extensions may be a good starting point or future investigations.

Even with these excellent data, these sources are not ideally modelled. Despite this, we believe the integrity of the key findings are intact and recommend additional study of these two systems. In particular for J1227, additional spectroscopy and radio interferometry would significantly improve the constraints on the distance and inclination.

\section*{Acknowledgements}
JGS, RPB, CJC, GV, MRK, and DMS acknowledge the support of the European Research Council, under the European Union's Horizon 2020 research and innovation program (grant agreement No. 715051; Spiders). VSD and TM acknowledge the support of the Science and Technology Facilities Council (STFC), grant number ST/T000406/1. TS thanks the Spanish Ministry of Economy and Competitiveness (MINECO; grant AYA2017-83216). This work has made use of data from the European Space Agency (ESA) mission {\it Gaia} (\url{https://www.cosmos.esa.int/gaia}), processed by the {\it Gaia} Data Processing and Analysis Consortium (DPAC, \url{https://www.cosmos.esa.int/web/gaia/dpac/consortium}). Funding for the DPAC has been provided by national institutions, in particular the institutions participating in the {\it Gaia} Multilateral Agreement.

Lastly, this work has made extensive use of the \texttt{python} language, including the \texttt{astropy} \citep{astropy}, \texttt{matplotlib} \citep{matplotlib}, \texttt{numpy} \citep{numpy}, and \texttt{scipy} \citep{scipy} packages.

\section*{Data availability}
The ULTRACAM light curves are available in Zenodo (\href{https://doi.org/10.5281/zenodo.4680510}{10.5281/zenodo.4680510}). The raw ULTRACAM images are available upon request. 

%%%%%%%%%%%%%%%%%%%%%%%%%%%%%%%%%%%%%%%%%%%%%%%%%%

%%%%%%%%%%%%%%%%%%%% REFERENCES %%%%%%%%%%%%%%%%%%

% The best way to enter references is to use BibTeX:

\bibliographystyle{mnras}
\bibliography{bib} % if your bibtex file is called example.bib

%%%%%%%%%%%%%%%%%%%%%%%%%%%%%%%%%%%%%%%%%%%%%%%%%%

%%%%%%%%%%%%%%%%% APPENDICES %%%%%%%%%%%%%%%%%%%%%

\appendix
\newcommand{\vnablapara}{\vec{\nabla}_\parallel}
\newcommand{\vJpara}{\vec{J}_\parallel}
\newcommand{\Tdh}{T_{\mathrm{dh}}}

\section{Icarus model extensions}\label{appendix:models}
\subsection{Hot spot model}
For the hot spot model, we consider the spot temperature, $T_{\rm spot}$, the spot radius, $R_{\rm spot}$, and the spot position angles, $\theta_s, \phi_s$. $\theta$ and $\phi$ are the polar angle and azimuth angle, such that $\theta_s = 0$\textdegree\ is the North pole of the companion and $\phi_s = 90$\textdegree\ is the direction towards the L1 point. The spot temperature is added to the base temperature of the companion after the effects of gravity darkening, but before the irradiation such that $T_{\rm star}(\theta_s,\phi_s)^4 = (T_{0} + T_{\rm spot}(\theta_s, \phi_s))^4 + T_{\rm irr}(\theta_s, \phi_s)^4$. In practice, there is little difference between this configuration and applying the spot after the irradiation; only the width of the spot would change. The spot geometry is defined by a 2D, axially symmetric Gaussian profile with a central maximum temperature of $T_{\rm spot}$ and width of $R_{\rm spot}$, with $R_{\rm spot} < 90$\textdegree.
 
\subsection{Heat redistribution model}
Energy transport follows the model \citep{Voisin2020}
\begin{equation}
\label{eq:nrjredist}
\vnablapara\cdot \vJpara = - \left(\sigma_{\rm sb}\left(T_*^4 - T_0^4\right) -  L_{\rm w}\right),
\end{equation}
which reduces to direct heating when the right-hand side is zero, and 
\begin{equation}
\label{eq:hr_law}
    \mathbf{J}_{\parallel} = -\kappa \left( \frac{T_*}{T_{\rm max}}\right)^{\Gamma} \nabla_{\parallel}T_{*} - T_* f(\theta) \sin(\theta) \mathbf{u}_{\phi},
\end{equation}
which is a generalisation of the parallel energy transport law derived in \citet{Voisin2020} (see appendix \ref{appendix:hrap}). $\mathbf{J}_{\parallel}$ is the surface energy flux, $\nabla_{\parallel}$ is the `surface gradient', $\kappa$ is the diffusion coefficient, $T_{*}$ is the surface temperature of the companion, $\Gamma$ is the diffusion index, $f(\theta)$ is the polar convection profile, and $\mathbf{u}_{\phi}$ is the unit vector of the longitude. $T_{\rm max}$ is an arbitrary constant which we define as $T_{\rm max}^4 = T_{\rm day}^4 = T_0^4 + T_{\rm irr}^4$. For this extension to the model, note that the spherical coordinates are defined differently to those in the hot spot model. $\theta$ is the colatitude, with the spherical coordinates defined such that the polar axis is the spin axis of the star, with $\phi = 0$ intersecting the binary axis on the night side of the star.

Out of several forms of polar convection profile, $f(\theta)$, we initially chose $f(\theta) = v$, where $v$ is the strength of the convection current in energy flux per unit temperature and is the first additional model parameter. This profile describes constant longitudinal advection. We also chose a convection profile of the form $f(\theta) = v \exp \left(- \frac{\theta^2}{2w^2}\right)$ which localises this flow to a Gaussian region around the equator with angular width $w$. We allow for the diffusion coefficient $\kappa \left(\frac{T_*}{T_{\rm max}}\right)^\Gamma$ to depend on the local temperature following a power law of index $\Gamma$.

\section{Heat redistribution with temperature-dependent diffusion law}
\label{appendix:hrap}
\subsection{Diffusion law and equation of energy redistribution}
In \citet{Voisin2020}, it is only described the case of the a constant diffusion coefficient $\kappa$. In the present paper, we make use of a conceptually straightforward generalisation of this work by introducing a temperature dependant diffusion coefficient whereby $\kappa \rightarrow \kappa(T) =  \kappa_\max (T/T_\max)^\Gamma$ as shown in Eq. \eqref{eq:hr_law}. For clarity and simplicity we use here $T\equiv T_*$, and the constant coefficient $\kappa$ of equation \eqref{eq:hr_law} is here written $\kappa_\max$.

In this case, $\kappa_\max$ is the value of the temperature-dependent coefficient at $T = T_{\max}$, and $\Gamma$ is the diffusion power-law index. Scaling the temperature by a ``maximum'' temperature $T_{\max}$ allows us to decorrelate the effect of variation of the diffusion index $\Gamma$ from variations of $\kappa_\max$. The choice of $T_{\max}$ is somewhat arbitrary (but should be such that $0 < T/T_{\max} \lesssim 1$ for the reason mentioned above). It is convenient to define 
\begin{equation}
T_{\max} = \left(T_0^4 + \bar T_{\rm ir}^4\right)^{1/4},
\end{equation}
where $T_0$ is the base temperature of the star, $L_w(a) = \sigma_{\rm sb} \bar T_{\rm ir}^4$, $L_w(a)$ being the wind flux at the separation distance $a$, and $\sigma_{\rm sb}$ the Stefan-Boltzmann constant. In the case of direct heating of a spherical star at homogeneous surface temperature by a spherically symmetric pulsar wind, this is indeed the maximum temperature at the surface of the companion star.

The equation of energy redistribution, Eq. \eqref{eq:nrjredist}, now becomes
\begin{equation}
\label{eq:nrjredist3}
\vnablapara\cdot\left(\kappa(T) \vnablapara T\right) + f(\theta)\partial_\phi T = \sigma_{\rm sb}\left(T^4 - T_0^4\right) - L_{\rm w},
\end{equation}
where the diffusion term can be expanded as
\begin{equation}
\label{eq:difterms}
	\vnablapara\cdot\left(\kappa(T) \vnablapara T\right) = \kappa_\max \left[ \left(\frac{T}{T_{\max}}\right)^\Gamma\vnablapara^2 T +  \left(\frac{T}{T_{\max}}\right)^{\Gamma-1}\frac{\left(\vnablapara T\right)^2}{T_{\max}}\right]
\end{equation}

\subsection{Linearisation and numerical solution}

In \citet{Voisin2020} is explained how the redistribution equation can be solved iteratively as the limit of the sequence
\begin{equation}
\label{eq:apBTn}
T_{n+1} = T_n + t_{n+1},
\end{equation}
where $t_{n+1}$ is the solution of equation \eqref{eq:nrjredist3} linearised with respect to $T_n$,
and $T_0 = T_{\rm dh} ; t_0 = 0$ initialise the sequence. 

In the present case, this scheme appears to diverge when $\Gamma \neq 0$ and $\kappa_{\max}$ is sufficiently large, typically when $t_1 \gtrsim T_b$. However, we found that a solution can be computed for any reasonable $\kappa_{\max}$ and $\Gamma$ by first computing the $\Gamma = 0$ solution (with the desired $\kappa_{\max}$) using the above scheme, and subsequently computing the sought solution by substituting $T_{\rm dh}$ by the $\Gamma=0$ solution as the initial condition of a new iteration.

The linearised equation \eqref{eq:nrjredist3} can be cast in the form
\begin{equation}
\label{eq:tn_new}
\left(A_n - B_n - f(\theta) \partial_\phi \right)t_{n+1} = S_{n},
\end{equation}
with
\begin{eqnarray}
A_n & = & 4 \sigma_{\rm sb} T_n^3 , \\
S_n & = & S_n^{\rm (dif)} + f(\theta) \partial_\phi T_n - \sigma(T_n^4 - T_0^4) + L_{\rm w}.
\end{eqnarray}

The difference with Eqs. (A10)-(A12) of \citet{Voisin2020} lies in the diffusion operator $B_n$ and source $S_n^{\rm (dif)}$. These are obtained by linearising Eq. \eqref{eq:difterms} around $T_n$,
\begin{eqnarray}
\vnablapara\cdot\left(\kappa(T_{n+1}) \vnablapara T_{n+1}\right) & = & B_n t_{n+1} + S_n^{\rm (dif)} + \bigcirc\left(t_{n+1}^2\right) 
\end{eqnarray}
where, 
\begin{eqnarray}
B_n & = & \kappa' T_n^{\Gamma-2}\left[T_n^2 \vnablapara^2  + 2\Gamma T_n \vnablapara T_n \cdot \vnablapara \right. \nonumber \\
& & \left.+ \Gamma T_n\vnablapara^2T_n + \Gamma (\Gamma-1)  \left(\vnablapara T_n\right)^2 \right], \\
S_n^{\rm (dif)} & = & \kappa' T_n^{\Gamma-1}\left[T_n \vnablapara^2 T_n + \Gamma \left(\vnablapara T_n\right)^2\right],
\end{eqnarray}
where for simplicity with have noted $\kappa' \equiv \kappa_{\max}/T_{\max}^k$.

As explained in \citet{Voisin2020}, Eq. \eqref{eq:tn_new} can be solved algebraically by decomposing $t_{n+1}$ on the basis of spherical harmonics. Numerically, this can be done using publicly available tools such as \texttt{shtools} \citep{wieczorek_shtools_2018} \footnote{\url{https://shtools.github.io/SHTOOLS/}}.

The solution is then obtained using spectral methods (e.g. \citep{bonazzola_spectral_1998}): in the space of spherical harmonics multiplication by a function and derivatives become matrix operators (truncated to the desired order). In particular, multiplication by a function $g(\theta,\phi)$ is given by a matrix of components 
\begin{equation}
	M_{ij}^{\times} = \sum_k g_k \mu_{ijk},
\end{equation}
where $g_k$ are the components of the spherical harmonics decomposition of $g$ and $\mu$ is such that 
\begin{equation}
	Y_{\beta}Y_{\gamma} = \sum_\alpha  Y_{\alpha}\mu_{\alpha\beta\gamma},
\end{equation}
where $Y_\gamma$ are the set of spherical harmonics.

With complex spherical harmonics, the derivative with respect to $\phi$ is given by 
\begin{equation}
	M_{ab}^{\partial_\phi} = i m_a \delta_{ab},
\end{equation}
where $\delta_{ab} = 1$ if $a=b$ and $0$ otherwise, $m_a$ is the spherical harmonic degree corresponding to index $a$ and $i^2 = -1$. 

In order to calculate the $\theta$ derivative we use the recurrence relation of associated Legendre functions \cite{olver_nist_2010},
\begin{equation}
	\frac{\mathrm{d}P_l^m}{\mathrm{d}\theta}(\cos\theta) = \frac{1}{\sin\theta}\left[(l+1-m) P_{l+1}^m(\cos\theta) - (l+1)\cos\theta P_l^m(\cos\theta)\right],
\end{equation}
in order to decompose numerically each $\partial_\theta Y_l^m$ onto the basis of spherical harmonics and thus obtain the matrix $M^{\partial_\theta}$.

It follows that, for example, $g(\theta,\phi)\partial_\phi$ translates into $M^{\times}M^{\partial_\phi}$ in spherical harmonic space. 

% Don't change these lines
\bsp	% typesetting comment
\label{lastpage}
\end{document}